\def\be{\begin{equation}}
\def\ee{\end{equation}}
\def\bea{\begin{eqnarray*}}
\def\eea{\end{eqnarray*}}
\newcommand{\degree}{\ensuremath{^\circ}}

\documentclass[useAMS,usenatbib]{mn2e}
\usepackage{subfigure}
\usepackage{graphicx}
\usepackage{flafter}
\usepackage{lscape}
\usepackage{amssymb}
\usepackage{ulem}
\begin{document}

\title[AzTEC 1.1 mm-wavelength imaging of GOODS-S - I.]{Deep 1.1\,mm-wavelength imaging of the GOODS-S field by AzTEC/ASTE - I. Source catalogue and number counts}


\author[K.S. Scott et al.]{
K.S.~Scott,$^{1,2}$ 
M.S.~Yun,$^1$ 
G.W.~Wilson,$^1$ 
J.E.~Austermann,$^3$
E. Aguilar,$^4$
\newauthor
I.~Aretxaga,$^4$ 
H.~Ezawa,$^5$ 
D.~Ferrusca,$^4$ 
B.~Hatsukade,$^6$ 
D.H.~Hughes,$^4$ 
D.~Iono,$^{5,6}$ 
\newauthor
M.~Giavalisco,$^1$
R.~Kawabe,$^5$ 
K.~Kohno,$^{6,7}$ 
P.D.~Mauskopf,$^8$ 
T.~Oshima,$^5$ 
\newauthor
T.A.~Perera,$^9$ 
J.~Rand,$^{1,10}$ 
Y.~Tamura,$^{5,11,12}$ 
T.~Tosaki,$^{13}$ 
M.~Velazquez,$^{4}$ 
\newauthor
C.C.~Williams,$^1$
and
M. Zeballos$^4$ \\
$^1$Department of Astronomy, University of Massachusetts, Amherst, MA 01003, USA\\
$^2$Department of Physics and Astronomy, University of Pennsylvania, Philadelphia, PA 19104, USA\\
$^3$Center for Astrophysics and Space Astronomy, University of Colorado, Boulder, CO 80309, USA\\
$^4$Instituto Nacional de Astrof\'{i}sica, \'{O}ptica y Electr\'{o}nica, Aptdo. Postal 51 y 216, 72000 Puebla, M\'{e}xico\\
$^5$Nobeyama Radio Observatory, National Astronomical Observatory of Japan, Minamimaki, Minamisaku, Nagano 384-1305, Japan\\
$^6$Institute of Astronomy, University of Tokyo, 2-21-1 Osawa, Mitaka, Tokyo 181-0015, Japan\\
$^7$Research Center for the Early Universe, University of Tokyo, 7-3-1 Hongo, Bunkyo, Tokyo 113-0033, Japan\\
$^8$School of Physics \& Astronomy, Cardiff University, Queens Buildings, The Parade, Cardiff CF24 3AA, UK\\
$^9$Department of Physics, Illinois Wesleyan University, Bloomington, IL 61701, USA\\
$^{10}$Department of Astronomy, University of Texas, Austin, TX 78712, USA\\
$^{11}$National Astronomical Observatory of Japan, 2-21-1 Osawa, Mitaka, Tokyo 181-8588, Japan\\
$^{12}$Department of Astronomy, University of Tokyo, Hongo, Bunkyo-ku, Tokyo 113-0033, Japan\\
$^{13}$Joetsu University of Education, 1 Yamayashiki-machi, Joetsu, Niigata, 943-8512, Japan\\
}
\date{\today}

\pagerange{\pageref{firstpage}--\pageref{lastpage}} \pubyear{2008}

\maketitle

\label{firstpage}


\begin{abstract}
We present the first results from a $1.1\,\rm{mm}$ confusion-limited map of the Great Observatories Origins Deep Survey-South (GOODS-S) taken with the AzTEC camera on the Atacama Submillimeter Telescope Experiment. We imaged a $270\,\rm{arcmin}^2$ field to a $1\sigma$ depth of $0.48-0.73\,\rm{mJy/beam}$, making this one of the deepest blank-field surveys at mm-wavelengths ever achieved. Although by traditional standards our GOODS-S map is extremely confused due to a sea of faint underlying sources, we demonstrate through simulations that our source identification and number counts analyses are robust, and the techniques discussed in this paper are relevant for other deeply confused surveys. We find a total of 41 dusty starburst galaxies with signal to noise ratios $S/N\ge3.5$ within this uniformly covered region, where only two are expected to be false detections, and an additional seven robust source candidates located in the noisier ($1\sigma\approx1\,\rm{mJy/beam}$) outer region of the map. We derive the $1.1\,\rm{mm}$ number counts from this field using two different methods: a fluctuation or ``$P(d)$'' analysis and a semi-Bayesian technique, and find that both methods give consistent results. Our data are well-fit by a Schechter function model with $(S^{\prime},N_{3\rm{mJy}},\alpha)=(1.30^{+0.19}_{-0.25}\,\rm{mJy},160^{+27}_{-28}\,\rm{mJy}^{-1}\rm{deg}^{-2},-2.0)$. Given the depth of this survey, we put the first tight constraints on the $1.1\,\rm{mm}$ number counts at $S_{1.1\rm{mm}}=0.5\,\rm{mJy}$, and we find evidence that the faint-end of the number counts at $S_{850\mu\rm{m}}\lesssim2.0\,\rm{mJy}$ from various SCUBA surveys towards lensing clusters are biased high. In contrast to the 870\,\micron~survey of this field with the LABOCA camera, we find no apparent under-density of sources compared to previous surveys at 1.1\,mm; the estimates of the number counts of SMGs at flux densities $\mathbf{>1}$\,mJy determined here are consistent with those measured from the AzTEC/SHADES survey. Additionally, we find a significant number of SMGs not identified in the LABOCA catalogue. We find that in contrast to observations at $\lambda\le500$\,\micron, MIPS 24\,\micron~sources do not resolve the total energy density in the cosmic infrared background at $1.1\,\rm{mm}$, demonstrating that a population of $z\gtrsim3$ dust-obscured galaxies that are unaccounted for at these shorter wavelengths potentially contribute to a large fraction {\bf ($\sim2/3$)} of the infrared background at $1.1\,\rm{mm}$.
\end{abstract}




\section{Introduction}
\label{sec:int} 

Galaxies selected at submillimetre (sub-mm) and millimetre (mm) wavelengths (hereafter SMGs) comprise a population of dust-obscured starburst or active galactic nuclei (AGN) host galaxies at high redshift \citep[$z\gtrsim1$; see review by][]{blain02}. With far-infrared (FIR) luminosities $L_{\rm{FIR}}\gtrsim10^{12}\,\rm{L}_\odot$, these systems appear to be scaled-up analogs to the ultra-luminous infrared galaxies (ULIRGs) observed in the local Universe \citep{sanders96}. Their FIR to mm spectral energy distributions (SEDs) are characterised by thermal dust emission with temperatures of $T_d\sim35-40$\,K \citep{chapman05, kovacs06, pope06, coppin08}, peaking in the FIR at rest-frame $\lambda\sim100$\,\micron. Due to the steep rise with frequency of the SED on the Rayleigh-Jeans tail \citep[$S_{\nu}\propto\nu^{3-4}$;][]{dunne00, dunne01}, the FIR peak is increasingly redshifted into the sub-mm/mm observing bands with increasing distance, resulting in a strong negative k-correction that roughly cancels the effects of cosmological dimming with redshift for observations at $\lambda\gtrsim500$\,\micron. This makes SMGs of a given bolometric luminosity equally detectable between $1<z<10$ at $1.1\,\rm{mm}$, assuming that a sufficient amount of dust can be built up in these systems at such early epochs. Extragalactic surveys at sub-mm/mm wavelengths have long taken advantage of this unique property and have detected hundreds of starburst galaxies in the early Universe, many of which go undetected in even the deepest optical surveys due to extreme dust-obscuration \citep[e.g.][]{younger07,younger09}. These include deep, large-area surveys towards ``blank-fields'', i.e. regions devoid of known galaxy over-densities \citep[e.g.][]{borys03, greve04, greve08, laurent05, coppin06, bertoldi07, scott08, perera08, weiss09, austermann10} and towards ``biased'' fields, such as the environments of high-redshift radio galaxies \citep[e.g.][]{stevens03} and other tracers of high-redshift proto-clusters \citep[e.g.][]{tamura09}. Several small-area surveys towards known clusters at moderate redshifts ($z\lesssim0.5$) have also been carried out in order to detect intrinsically less luminous background SMGs via amplification through gravitational lensing \citep[e.g.][]{smail98, smail02, chapman02, cowie02, knudsen06, knudsen08}.

The number counts at sub-mm/mm wavelengths provide potentially strong constraints on models of galaxy evolution \citep[e.g.][]{granato04, silva05, baugh05, rowan-robinson09}. The hundreds of SMGs detected over the past 12 years in large-area surveys taken with the Submillimeter Common-User Bolometer Array \citep[SCUBA, 850\,\micron;][]{holland99} on the 15-m James Clerk Maxwell Telescope (JCMT), MAMBO \citep[1.2\,mm;][]{kreysa98} on the Institut de Radio Astronomie Millimetrique (IRAM) 30-m telescope, Bolocam \citep[1.1\,mm;][]{glenn98, haig04} on the 10-m Caltech Submillimeter Observatory (CSO), and AzTEC \citep[1.1\,mm;][]{wilson08a} on the JCMT and the 10-m Atacama Submillimeter Telescope Experiment \citep[ASTE,][]{ezawa04,ezawa08} have put tight constraints on the number counts at $S_{850\mu\rm{m}}\ge2\,\rm{mJy}$ ($S_{1.1\rm{mm}}\ge1\,\rm{mJy}$). Regardless of the precise details of various galaxy evolution models, the observed number counts at sub-mm/mm wavelengths require strong luminosity evolution of IR-bright galaxies \citep[e.g.][]{scott02, scott06, greve05, coppin06, austermann10}. The low angular resolution (full width at half maximum $\rm{FWHM}\ge10$\arcsec) of these observations makes them confusion-limited at $\sim1\,\rm{mJy}$ (at both 850\,\micron~and $1.1\,\rm{mm}$ on the 15-m JCMT), limiting source detections to only the most luminous ($L_{\rm{FIR}}\gtrsim3\times10^{12}\,\rm{L}_\odot$) systems with extreme star-formation rates of $\rm{SFR}\gtrsim500\,\rm{M}_\odot/\rm{yr}$. Only surveys towards lensing clusters have picked out individual sources with lower luminosities, and these currently provide only weak constraints on the sub-mm number counts at $S_{850\mu\rm{m}}\lesssim2\,\rm{mJy}$. While detecting a large number of intrinsically fainter systems and establishing the link between SMGs and other high-redshift galaxy populations must await the improved resolution of the 50-m Large Millimeter Telescope (LMT) and the Atacama Large Millimeter/Submillimeter Array (ALMA), information about their number counts at $S_{1.1\rm{mm}}<1\,\rm{mJy}$ can be gleaned from wide-area, confusion-limited surveys with existing facilities, and these can aid in discriminating between various galaxy evolution models.

In this paper we present a $270\,\rm{arcmin}^2$ $1.1\,\rm{mm}$ survey of the Great Observatories Origins Deep Survey-South (GOODS-S) field taken with the AzTEC camera on the ASTE. This is the deepest survey at mm wavelengths ever carried out, achieving a root-mean-square (rms) noise level of $1\sigma=0.48-0.73\,\rm{mJy}$. The GOODS-S field represents one of the most widely observed regions of sky, with deep multi-wavelength data from a number of ground-based and space-based facilities. This includes X-ray data from \textit{Chandra} \citep{luo08}, optical to near-IR photometry from the \textit{Hubble Space Telescope} \citep[\textit{HST};][]{giavalisco04a}, near-IR imaging with ISAAC on the Very Large Telescope \citep{retzlaff09}, \textit{Spitzer} IRAC (Chary et al. in prep.) and MIPS (Dickinson et al. in prep.) imaging in the mid-IR, sub-mm imaging at $250$, $350$, and $500$\,\micron~with the Balloon-borne Large Aperture Submillimeter Telescope \citep[BLAST;][]{devlin09} and at 870\,\micron~with the LABOCA camera on the Atacama Pathfinder EXperiment \citep[APEX;][]{weiss09}, and $1.4\,\rm{GHz}$ interferometric imaging with the Very Large Array \citep[VLA;][]{kellermann08,miller08}. Dedicated spectroscopic follow-up of optical sources in this field has also been underway \citep{vanzella05, vanzella06, vanzella08, popesso09}. Additionally, planned \textit{Herschel} observations of the GOODS-S field will provide ultra-deep FIR/sub-mm imaging at 100, 250, 350, and 450~\micron. This suite of multi-wavelength data is essential for the identification of counterparts to the SMGs and for the characterisation of the properties of these galaxies.

This paper is organised as follows: in Section~\ref{sec:obs} we describe the observations of the GOODS-S field carried out with AzTEC on ASTE. In Section~\ref{sec:dat} we summarise the data reduction methods. We present the $1.1\,\rm{mm}$ map and source catalogue in Section~\ref{sec:scat}, and we describe simulations carried out to characterise the number of false detections, survey completeness, and degree of source blending in the map in Section~\ref{sec:confusion}. We derive the $1.1\,\rm{mm}$ number counts from this survey in Section~\ref{sec:nc} and compare them with the number counts determined from SCUBA lensing cluster surveys and existing blank-field surveys at $1.1\,\rm{mm}$ wavelengths. We present a comparison between the AzTEC/GOODS-S data and the 870\,\micron~data from LABOCA in Section~\ref{sec:laboca}. We discuss the contribution to the cosmic infrared background (CIRB) at $1.1\,\rm{mm}$ from the radio and mid-IR galaxy populations in Section~\ref{sec:cib}, and we close with a summary of our results in Section~\ref{sec:con}. Several upcoming papers involving this data-set are underway, including the identification of radio and mid-IR counterparts to the SMGs (Scott et al. in prep.), a study of the $1.1\,\rm{mm}$ properties of $BzK$-selected galaxies (Welch et al. in prep.), a comparison between the AzTEC/ASTE GOODS-S data and the sub-mm maps from BLAST (Aguilar et al. in prep.), and a study of the X-ray properties of SMGs (Johnson et al. in prep.).


\section{Observations}
\label{sec:obs}

We imaged a $26\times20\,\rm{arcmin}^2$ field centred at right ascension and declination (RA, Dec) $=$ (03$^{\rm h}$32$^{\rm m}$30$^{\rm s}$, $-$27\degree48\arcmin20\arcsec) at $1.1\,\rm{mm}$ using AzTEC on the ASTE. The central $19\times14\,\rm{arcmin}^2$ region, where the coverage is uniform, encompasses the entire GOODS-S region mapped by the \textit{Spitzer} Space Telescope. The observations were carried out using the N-COSMOS3 network observation system \citep{kamazaki05} from July 15 to August 6 during the 2007 Chilean winter under excellent observing conditions, with $\tau_{220}=0.05$ on average, and $\tau_{220}<0.06$ 70\% of the time (zenith opacity at $220\,\rm{GHz}$ reported by the ASTE tau monitor). A total of 52 hours of observing time excluding pointing and calibration overheads was devoted to this survey. During the 2007 season, $107$ (out of 144) of the AzTEC bolometers were stable with high sensitivity. The point spread function (PSF) of each detector was measured via beam-maps on Uranus, Neptune, and 3C279 as described in \citet{wilson08a} and has a FWHM of $30\arcsec\pm1\arcsec$ and $31\arcsec\pm2\arcsec$ in azimuth and elevation, respectively (the theoretical beam size is 27\arcsec~FWHM). The full array subtends a circular field-of-view with a diameter of $8\arcmin$.

\subsection{Scan Strategy}
\label{ssec:lissajous}

We used a continuous scanning strategy which traces a modified Lissajous pattern on the sky:
\begin{eqnarray}
\label{equ:lissajous}
\Delta \mbox{RA} & = & 5.5\arcmin \cdot \mbox{sin}(a \cdot t + 0.25) + 2.0\arcmin \cdot \mbox{sin}(a \cdot t/30) \nonumber\\
\Delta \mbox{Dec} & = & 7.5\arcmin \cdot \mbox{sin}(b \cdot t) + 2.0\arcmin \cdot \mbox{sin}(b \cdot t/30),
\end{eqnarray}
where $a/b=8/9$ and $\Delta$RA and $\Delta$Dec are physical coordinates relative to the field centre. The actual values of $a$ and $b$ were scaled to limit the peak scanning velocity to $300\arcsec/\rm{s}$, and a rotational angle $20\degree$ West of North was used in order to align our map with that of the \textit{Spitzer} IRAC/MIPS coverage of GOODS-S. A single observation took 42 minutes to complete, and we obtained a total of 74 observations of the GOODS-S field.

The benefit of Lissajous scanning, in addition to attaining excellent cross-linking and uniform coverage in the map, is that we avoid large telescope accelerations that can induce systematics in the detector signals as well as compromise the pointing accuracy of the telescope. Such effects are often seen in images taken in raster-scan mode, where $1/3-1/2$ of the data taken during times when the telescope reverses direction must be discarded \citep[e.g.][]{scott08,perera08}. Lissajous scanning, on the other hand, results in nearly 100\% observing efficiency.

\subsection{Pointing Corrections}
\label{ssec:pointing}

We make small corrections to the telescope pointing model by routinely observing the bright point source J0455$-$462 ($S_{1.1\rm{mm}}\sim1.5\,\rm{Jy}$, variable) every two hours before and after each block of GOODS-S observations. We measure pointing offsets by fitting the $4\times4\,\rm{arcmin}^2$ maps of J0455$-$462 to 2-dimensional Gaussians, and we linearly interpolate these offset corrections temporally and apply them to the GOODS-S data. The random pointing error in the final GOODS-S map is $\lesssim1\arcsec$ (see Section \ref{ssec:astrometry}).

\subsection{Flux Calibration}
\label{ssec:calibration}

The flux conversion factor (FCF) used to convert the raw detector signals to flux density units was determined by beam map observations on Uranus, Neptune, and 3C279, taken $1-2$ times per night as described in \citet{wilson08a}. The flux densities of Uranus and Neptune at $1.1\,\rm{mm}$ were calculated from their frequency-dependent brightness temperatures reported in \citet{griffin93} and ranged from $43-52\,\rm{Jy}$ and $18-20\,\rm{Jy}$, respectively, during the AzTEC/ASTE 2007 observing season. The flux density of 3C279, which is highly variable, is adopted from the Submillimeter Array flux density archive\footnote{http://sma1.sma.hawaii.edu/callist/callist.html} and ranged from $7.0-9.4\,\rm{Jy}$ at $1.1\,\rm{mm}$ during this time. We remove the responsivity factor from the detector signals and correct for extinction by modelling both as a linear function of the demodulated DC-level \citep[see][]{wilson08a}.

The measured FCF varied significantly from night to night, resulting in a $1\sigma$ scatter of 17\% over the entire observing run. We have identified the source of this scatter as the changing focal point of the telescope with environment temperature: the FCF decreases as the measured FWHM of the beam increases. Since real time corrections to the subreflector position were not possible, we use the same-night measurement of the FCF to calibrate each observation.  To estimate the total calibration uncertainty, we determine the standard deviation in the measured flux densities from the 68 pointing observations of J0455$-$462, which is 8\%. Since this source is known to be variable, this gives a conservative upper limit to our calibration uncertainty. Combining this in quadrature with the 5\% absolute uncertainty on the flux densities of Uranus and Neptune \citep{griffin93} gives a total upper limit to the calibration error of 9\%.


\section{Data Reduction}
\label{sec:dat}

We reduce the $1.1\,\rm{mm}$ data in a manner that is nearly identical to that described in detail in \citet{scott08}. We summarise the steps here and note the differences. The raw time-stream data, which consist of all bolometer signals and pointing data stored as a function of time, is first scanned for ``spikes'' (defined as any $>7\sigma$ jump between sequential detector samples), which are caused by instrumental glitches or cosmic ray strikes. These data and nearby samples, which amount to $<0.1$\% of the total time-stream data, are flagged and discarded from the data-set. We group the remaining samples into 10-second intervals, and then ``clean'' each 10-second group using a principal component analysis (PCA) to identify and remove the common-mode atmospheric signal \citep{laurent05,scott08}.

For AzTEC maps taken in raster-scan mode \citep[e.g.][]{scott08,perera08} data samples from the same individual scan (a single pass of the telescope across the sky) were grouped together for PCA cleaning. Since there is no such natural division for continuous Lissajous-scan maps, we chose a 10-second interval grouping. \citet{perera08} showed via a statistical correlation analysis that this PCA cleaning technique using time intervals ranging from $5-15$~seconds provides a good balance between using a sufficient number of samples to determine the bolometer-bolometer correlations and being on a short enough time scale so that the slower electronics related low-frequency drifts can be effectively removed. We have verified that cleaning on $5-20$-second intervals gives consistent results for AzTEC/GOODS-S.

After cleaning the time-stream data, the bolometer signals are calibrated and binned into $3\arcsec\times3\arcsec$ pixels to make a map for each separate observation. These 74 maps are then co-added and wiener filtered to suppress correlated large-scale structure from residual atmosphere and pixel-to-pixel variations on scales smaller than the beam in order to optimise the map for point source detection \citep{scott08}. In addition to this filtered map, we track the effects of PCA cleaning and filtering on a model point source (referred to hereafter as the point source kernel). The smoothing slightly broadens the FWHM of the beam: fitting a 2-dimensional Gaussian to the point source kernel results in a FWHM of $34.6\arcsec$ and $34.3\arcsec$ in RA and Dec, respectively. We also generate 100 noise maps, each a realisation of the signal-free noise in the GOODS-S map, by ``jackknifing'' the time-stream signals in the same manner described in \citet{scott08}. The point source kernel and noise realisations are used later in the analysis for simulation purposes (Section~\ref{sec:confusion}) and number counts determination (Section~\ref{sec:nc}).

\section{1.1\,mm Map and Source Catalogue}
\label{sec:scat}

\subsection{AzTEC 1.1\,mm Map}
\label{ssec:map}

\begin{figure*}
\begin{center}
\includegraphics{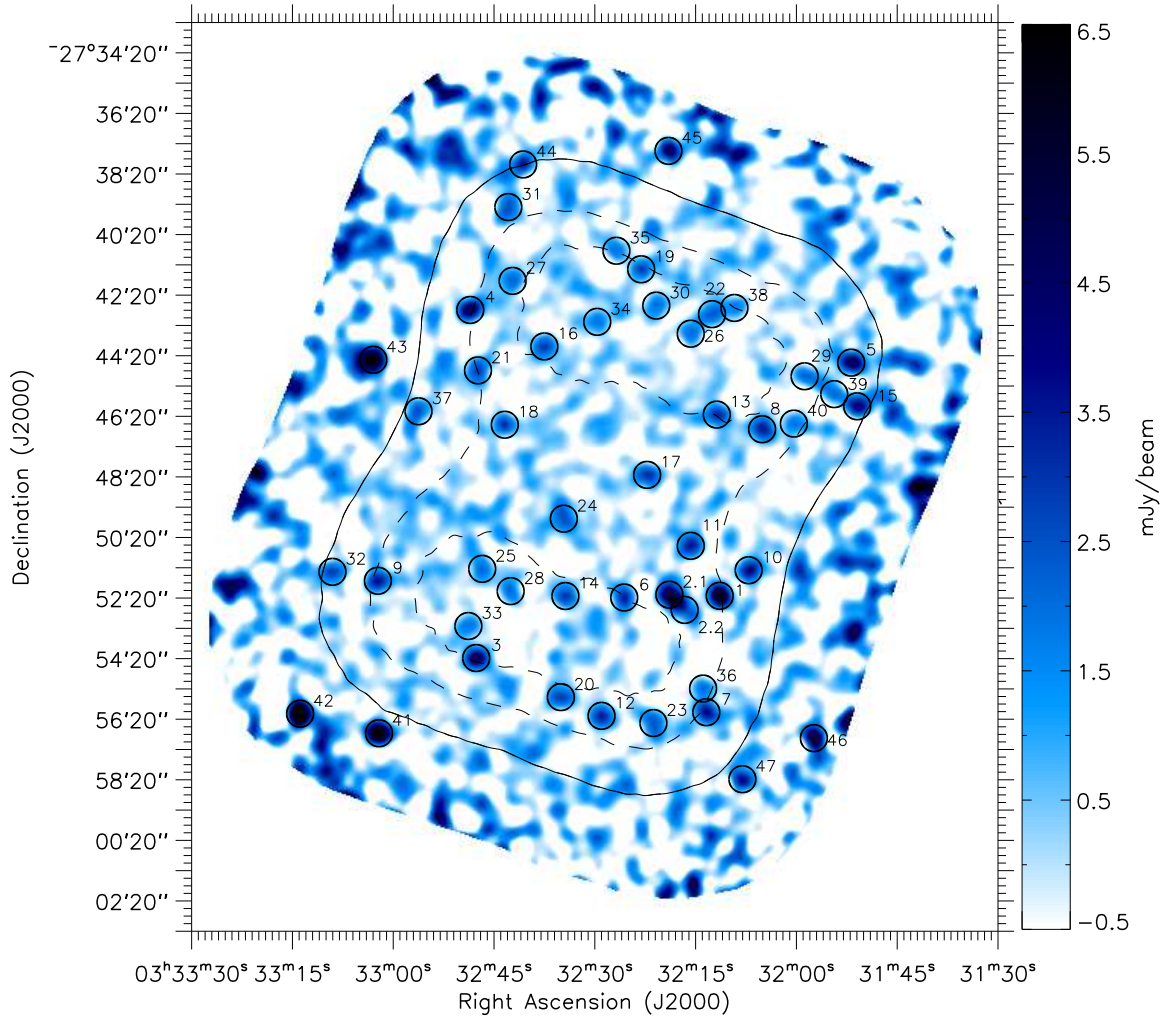}
\caption{The AzTEC $1.1\,\rm{mm}$ map of the GOODS-S field. The solid contour shows the boundary of the $270\,\rm{arcmin}^2$ 50\% uniform coverage region to which most of the analysis of this field is restricted. The dashed contours (innermost to outermost) indicate noise rms levels of $0.51$ and $0.57\,\rm{mJy/beam}$. The circles located within the solid contour (diameter $=2\times\rm{FWHM}$ of AzTEC on ASTE $=60\arcsec$) indicate the positions of $\ge3.5\sigma$ source candidates, labeled in order of decreasing $S/N$ of the detections. The circles located outside of the solid contour indicate robust $\ge4.5\sigma$ sources within the noiser regions of the map. See the on-line journal for a colour version of this Figure.}
\label{fig:survey_map} 
\end{center}
\end{figure*}

The full $512\,\rm{arcmin}^2$ $1.1\,\rm{mm}$ map of GOODS-S taken with AzTEC on ASTE is shown in Figure~\ref{fig:survey_map}. Since we must consider only a region with uniform noise for most of the analysis presented in this paper, we define a $270\,\rm{arcmin}^2$ ``uniform coverage region'', where the coverage is greater than or equal to 50\% of the maximum coverage in the map. The $1\sigma$ rms noise in this region ranges from $0.48-0.73\,\rm{mJy/beam}$, making this the deepest contiguous region ever mapped at $1.1\,\rm{mm}$. The scanning strategy that we used to map this field results in two separate patches where the coverage is deepest ($0.48\,\rm{mJy/beam}$), whereas the center of the map is slightly shallower ($0.56\,\rm{mJy/beam}$, see rms contours on Figure \ref{fig:survey_map}). The noise determined from the jackknifed noise realizations is Gaussian distributed with $1\sigma=0.57$\,mJy/beam.

\subsection{Astrometry}
\label{ssec:astrometry}

To check for a residual astrometric offset after applying the pointing model (see Section~\ref{ssec:pointing}) we stack the $1.1\,\rm{mm}$ GOODS-S map at the positions of radio sources in this field, whose positions are known to much better than $1\arcsec$. From the VLA $1.4\,\rm{GHz}$ survey of the Extended Chandra Deep Field-South \citep[ECDF-S;][]{miller08}, which reaches an rms noise level of $7-10\,\mu\rm{Jy}$, we extract a $\ge4\sigma$ radio source catalogue\footnote{The radio source catalogue is produced using the SAD program in the Astronomical Image Processing System (AIPS; http://www.aips.nrao.edu/), which uses 2-dimensional Gaussian modelling to identify sources and produce photometry results.} and stack the AzTEC map at the positions of the 219 radio sources that lie within the 50\% uniform coverage region.  The resulting stacked AzTEC map shows a clear peak with signal to noise $S/N=8$, and its centroid is offset by $-6.3\arcsec\pm2.1\arcsec$ in RA and $-3.9\arcsec\pm2.1\arcsec$ in Dec. We have verified this result by stacking at the locations of 1185 MIPS 24~\micron~sources detected with $S/N>10$ in the \textit{Spitzer} GOODS-S survey (Dickinson et al. in prep.), which gives a $S/N=11$ detection with a centroid offset of $-6.3\arcsec\pm1.6\arcsec$ in RA and $-0.7\arcsec\pm1.6\arcsec$ in Dec, consistent with the radio stacking results. We thus apply an astrometric correction of $(\Delta\rm{RA}, \Delta\rm{Dec}) = (6.3\arcsec, 0.7\arcsec)$ to the AzTEC map and $1.1\,\rm{mm}$ source positions (we favor the offsets measured from the 24~\micron~stacking due to the higher $S/N$ of the peak detection).

This systematic offset represents the \textit{average} pointing offset between J0455$-$462 and the GOODS-S field over 74 observations. The scatter in this offset, or the pointing jitter, will manifest itself as a broadening of the source in excess of the point source kernel. We estimate the random pointing error in the AzTEC map from the brightest AzTEC source in this field, AzTEC/GS1. We adopt a simple model which consists of the convolution of the ideal point source kernel with a 2-dimensional Gaussian with standard deviation $(\sigma_{\rm{RA}},\sigma_{\rm{Dec}})$, where $\sigma_{\rm{RA}}$ and $\sigma_{\rm{Dec}}$ are the $1\sigma$ random pointing errors in RA and Dec \citep[see][for a full description of this measurement]{scott08}. The high $S/N$ of AzTEC/GS1 ($11.6\sigma$, see Table~\ref{table:sources}) allows a clean measurement and provides a strong constraint on the random pointing error, since the signal from this single source can only be broadened with respect to the point source kernel due to the scatter in the pointing model. The maximum likelihood estimate for the random pointing error from AzTEC/GS1 is $(\sigma_{\rm{RA}},\sigma_{\rm{Dec}}) = (0.5\arcsec,0.1\arcsec)$; however, the distribution of $(\sigma_{\rm{RA}},\sigma_{\rm{Dec}})$ is very flat out to $1\arcsec$, then falls off steadily. From this we conclude that the random pointing error in the AzTEC/GOODS-S map is $\lesssim1\arcsec$.

While the stacked $1.1\,\rm{mm}$ signal for the radio and MIPS 24~\micron~populations is valuable for determining the astrometric offset, we find that the stacked signal cannot be used for determining the pointing jitter as it is significantly broader than the point source kernel, with $1\sigma\approx6.8\arcsec$ and $15\arcsec$ for the stack on the radio and 24~\micron~sources, respectively. This additional broadening is likely caused by other effects such as confused/blended sources in the $1.1\,\rm{mm}$ image or clustering of the radio and 24~\micron~sources.

\subsection{Source Catalogue}
\label{ssec:sources}

We identify point sources in the $1.1\,\rm{mm}$ $S/N$ map by searching for local maxima within $15\arcsec$ of pixels with $S/N\ge3.5$ \citep[see][for a more detailed description]{scott08}. The 40 source candidates located within the uniform coverage region that meet this criterion are listed in Table~\ref{table:sources} in order of decreasing $S/N$ of the detection. We note that the number of selected source candidates is independent of the window size for grouping high $S/N$ pixels, for values ranging from 3\arcsec to 45\arcsec. Table~\ref{table:sources} includes  both the $1.1\,\rm{mm}$ flux densities and $1\sigma$ errors measured from the map, as well as the bias-corrected flux densities estimated using a semi-Bayesian technique \citep[Section~\ref{ssec:bayes};][]{coppin05, coppin06, austermann09, austermann10}. This flux-bias correction accounts for the fact that the measured flux densities of mm-selected galaxies, which are generally detected at low $S/N$, are preferentially ``boosted'' due to the steep luminosity distribution of the population \citep[e.g.][]{hogg98}.

We find an additional seven robust source candidates that are located outside of the uniform coverage region, but are detected with high $S/N$ ($\ge4.5$). These sources are listed in Table~\ref{table:sources} for the benefit of future studies; however, we do not address the general properties of these sources located outside of the uniform coverage region (e.g. number of false detections, completeness) nor include them in our number counts estimation in Section~\ref{sec:nc}.

Several of the source candidates appear extended in the $1.1\,\rm{mm}$ map, most notably, AzTEC/GS2 (labelled as 2.1 and 2.2 in Figure \ref{fig:survey_map}). To separate the components of AzTEC/GS2, we fit the $1.1\,\rm{mm}$ map in the neighbourhood of this source to a 2-component model, where each component is a scaled version of the point source kernel. The best-fit positions and flux densities are listed in Table \ref{table:sources}. Given the comparatively modest $S/N$ of the other AzTEC sources which are potentially extended, we cannot fit a model to these sources in order to separate them into multiple components.

\begin{table*}
\caption{\label{table:sources}AzTEC/GOODS-S source candidates. The columns give: 1) AzTEC source identification; 2) source name; 3) $S/N$ of the detection in the AzTEC map; 4) measured 1.1\,mm flux density and error; and 5) deboosted 1.1\,mm flux density and 68\% confidence interval (Section~\ref{ssec:bayes}). The sources above the horizontal line are detected with $S/N\ge3.5$ within the uniform coverage region of the AzTEC/GOODS-S map, where two are expected to be false positives. The $S/N\ge4.5$ sources listed below the horizontal line are located outside the uniform coverage region.}
\begin{center}
\begin{tabular}{llrcc}
\hline
          &               &                             &  $S_{1.1\rm{mm}}$  & $S_{1.1\rm{mm}}$ \\
          &               &                             &  (measured)        & (deboosted)      \\
AzTEC ID  &  Source Name  &  \multicolumn{1}{c}{$S/N$}  &  (mJy)             & (mJy)            \\
\hline
AzTEC/GS1$^*$ & AzTEC\_J033211.46-275216.0 & 11.6 & $6.6\pm0.6$ & $6.3^{+0.5}_{-0.6}$               \\
AzTEC/GS2 & AzTEC\_J033218.48-275221.8 & 11.4 & $6.0\pm0.5$ & $5.7^{+0.5}_{-0.6}$               \\
\hspace*{1.15cm}GS2.1$^*$ & AzTEC\_J033218.99-275213.8 &      & $6.6\pm0.5$ & $6.3^{+0.5}_{-0.5}$   \\
\hspace*{1.15cm}GS2.2$^*$ & AzTEC\_J033216.70-275244.0 &      & $4.0\pm0.5$ & $3.7^{+0.5}_{-0.5}$   \\
AzTEC/GS3$^*$ & AzTEC\_J033247.86-275419.3 &  9.4 & $4.8\pm0.5$ & $4.5^{+0.5}_{-0.5}$               \\
AzTEC/GS4$^*$ & AzTEC\_J033248.75-274249.5 &  8.6 & $5.0\pm0.6$ & $4.6^{+0.6}_{-0.6}$               \\
AzTEC/GS5$^*$ & AzTEC\_J033151.81-274433.9 &  7.8 & $4.8\pm0.6$ & $4.4^{+0.6}_{-0.6}$               \\
AzTEC/GS6$^*$ & AzTEC\_J033225.73-275219.4 &  6.7 & $3.4\pm0.5$ & $3.1^{+0.5}_{-0.5}$               \\
AzTEC/GS7$^*$ & AzTEC\_J033213.47-275606.7 &  6.7 & $3.9\pm0.6$ & $3.5^{+0.6}_{-0.6}$               \\
AzTEC/GS8$^*$ & AzTEC\_J033205.12-274645.8 &  6.6 & $3.5\pm0.5$ & $3.1^{+0.5}_{-0.5}$               \\
AzTEC/GS9 & AzTEC\_J033302.56-275146.1 &  6.5 & $3.6\pm0.6$ & $3.2^{+0.6}_{-0.5}$               \\
AzTEC/GS10$^*$ & AzTEC\_J033207.13-275125.3 &  6.3 & $3.9\pm0.6$ & $3.5^{+0.6}_{-0.6}$              \\
AzTEC/GS11 & AzTEC\_J033215.79-275036.8 &  6.2 & $3.4\pm0.6$ & $3.1^{+0.6}_{-0.6}$              \\
AzTEC/GS12$^*$ & AzTEC\_J033229.13-275613.8 &  6.2 & $3.3\pm0.5$ & $2.9^{+0.5}_{-0.5}$              \\
AzTEC/GS13 & AzTEC\_J033211.91-274616.9 &  6.2 & $3.1\pm0.5$ & $2.8^{+0.5}_{-0.5}$              \\
AzTEC/GS14 & AzTEC\_J033234.52-275216.4 &  6.0 & $3.0\pm0.5$ & $2.6^{+0.5}_{-0.5}$              \\
AzTEC/GS15$^*$ & AzTEC\_J033150.91-274600.4 &  6.0 & $4.0\pm0.7$ & $3.5^{+0.7}_{-0.7}$              \\
AzTEC/GS16 & AzTEC\_J033237.67-274401.8 &  5.7 & $2.8\pm0.5$ & $2.5^{+0.5}_{-0.5}$              \\
AzTEC/GS17 & AzTEC\_J033222.31-274816.4 &  5.6 & $3.1\pm0.6$ & $2.7^{+0.5}_{-0.6}$              \\
AzTEC/GS18$^*$ & AzTEC\_J033243.58-274636.9 &  5.5 & $3.1\pm0.6$ & $2.7^{+0.5}_{-0.6}$              \\
AzTEC/GS19 & AzTEC\_J033223.21-274128.8 &  5.4 & $2.7\pm0.5$ & $2.4^{+0.5}_{-0.5}$              \\
AzTEC/GS20 & AzTEC\_J033235.22-275536.8 &  5.2 & $2.7\pm0.5$ & $2.4^{+0.5}_{-0.5}$              \\
AzTEC/GS21 & AzTEC\_J033247.60-274449.3 &  5.0 & $2.9\pm0.6$ & $2.4^{+0.6}_{-0.6}$              \\
AzTEC/GS22 & AzTEC\_J033212.60-274257.9 &  4.7 & $2.3\pm0.5$ & $2.0^{+0.5}_{-0.5}$              \\
AzTEC/GS23$^*$ & AzTEC\_J033221.37-275628.1 &  4.7 & $2.5\pm0.5$ & $2.1^{+0.6}_{-0.5}$              \\
AzTEC/GS24 & AzTEC\_J033234.76-274943.1 &  4.6 & $2.5\pm0.5$ & $2.1^{+0.6}_{-0.6}$              \\
AzTEC/GS25$^*$ & AzTEC\_J033246.96-275122.4 &  4.4 & $2.2\pm0.5$ & $1.8^{+0.5}_{-0.5}$              \\
AzTEC/GS26 & AzTEC\_J033215.79-274336.6 &  4.4 & $2.1\pm0.5$ & $1.8^{+0.5}_{-0.5}$              \\
AzTEC/GS27 & AzTEC\_J033242.42-274151.9 &  4.3 & $2.3\pm0.5$ & $1.8^{+0.5}_{-0.5}$              \\
AzTEC/GS28 & AzTEC\_J033242.71-275206.8 &  4.3 & $2.1\pm0.5$ & $1.7^{+0.5}_{-0.5}$              \\
AzTEC/GS29 & AzTEC\_J033158.77-274500.9 &  4.1 & $2.2\pm0.5$ & $1.8^{+0.5}_{-0.6}$              \\
AzTEC/GS30 & AzTEC\_J033220.94-274240.8 &  4.1 & $2.0\pm0.5$ & $1.7^{+0.5}_{-0.5}$              \\
AzTEC/GS31 & AzTEC\_J033243.06-273925.6 &  4.1 & $2.5\pm0.6$ & $2.0^{+0.6}_{-0.7}$              \\
AzTEC/GS32 & AzTEC\_J033309.35-275128.4 &  4.1 & $2.8\pm0.7$ & $2.1^{+0.7}_{-0.7}$              \\
AzTEC/GS33 & AzTEC\_J033249.03-275315.8 &  4.1 & $2.0\pm0.5$ & $1.6^{+0.5}_{-0.5}$              \\
AzTEC/GS34 & AzTEC\_J033229.77-274313.1 &  4.0 & $2.0\pm0.5$ & $1.6^{+0.5}_{-0.5}$              \\
AzTEC/GS35 & AzTEC\_J033226.90-274052.1 &  4.0 & $2.0\pm0.5$ & $1.6^{+0.5}_{-0.5}$              \\
AzTEC/GS36 & AzTEC\_J033213.94-275519.7 &  3.7 & $2.1\pm0.6$ & $1.5^{+0.6}_{-0.5}$              \\
AzTEC/GS37 & AzTEC\_J033256.48-274610.3 &  3.7 & $2.5\pm0.7$ & $1.8^{+0.7}_{-0.7}$              \\
AzTEC/GS38 & AzTEC\_J033209.26-274245.5 &  3.6 & $1.8\pm0.5$ & $1.4^{+0.5}_{-0.5}$              \\
AzTEC/GS39$^*$ & AzTEC\_J033154.34-274536.3 &  3.5 & $2.1\pm0.6$ & $1.5^{+0.6}_{-0.6}$              \\
AzTEC/GS40 & AzTEC\_J033200.38-274634.6 &  3.5 & $1.9\pm0.6$ & $1.4^{+0.6}_{-0.6}$              \\
\hline
AzTEC/GS41$^*$ & AzTEC\_J033302.39-275648.4 &  8.0 & $7.1\pm0.9$ & $6.3^{+0.9}_{-0.9}$              \\
AzTEC/GS42$^*$ & AzTEC\_J033314.19-275609.6 &  7.8 & $9.2\pm1.2$ & $7.9^{+1.1}_{-1.4}$              \\
AzTEC/GS43$^*$ & AzTEC\_J033303.24-274428.3 &  6.9 & $6.7\pm1.0$ & $5.7^{+1.0}_{-1.0}$              \\
AzTEC/GS44$^*$ & AzTEC\_J033240.84-273801.1 &  5.0 & $3.7\pm0.7$ & $3.0^{+0.8}_{-0.8}$              \\
AzTEC/GS45$^*$ & AzTEC\_J033219.12-273734.1 &  4.8 & $4.8\pm1.0$ & $3.6^{+1.0}_{-1.1}$              \\
AzTEC/GS46$^*$ & AzTEC\_J033157.38-275658.0 &  4.7 & $5.9\pm1.2$ & $4.0^{+1.3}_{-1.4}$              \\
AzTEC/GS47$^*$ & AzTEC\_J033208.06-275819.4 &  4.5 & $3.8\pm0.8$ & $2.8^{+0.9}_{-0.8}$              \\
\hline
\end{tabular}
\end{center}
$^*$ These source candidates have probable $870\,\micron$ counterparts detected with $S/N\ge3.7$ in the LABOCA map (Section \ref{ssec:aztec_laboca_smatch}).
\end{table*}

\section{Characterisation of a Deeply Confused Map}
\label{sec:confusion}

Source confusion arising from a significant underlying population of faint sources that are individually undetected in a given survey is a function of source density and resolution \citep[e.g.][and references therein]{hogg01}, and is important to consider for our AzTEC/GOODS-S survey. Using the standard rule of thumb (one source per 30 beams)\footnote{By deriving a general formalisation of source confusion statistics, \citet{takeuchi04} demonstrate the basis for the ``one source per 30 beams'' rule of thumb for estimating the confusion limit. \citet{takeuchi04} also show that the actual confusion limit is strongly dependant on the steepness of the source counts, rather than simply the source density.} and the blank-field $1.1\,\rm{mm}$ source counts from the AzTEC/SHADES survey \citep{austermann10}, the confusion limit given the $30\arcsec$ FWHM ASTE beam is $2.0\,\rm{mJy}$. Since our AzTEC/GOODS-S survey is $3-4$ times deeper than the formal confusion limit, we use specialised simulations to characterise the effects of source confusion on the properties of our GOODS-S map, including the number of false detections, the survey completeness, the positional uncertainty distribution, and the degree of source blending.

For these purposes, we generate sky realisations of the GOODS-S field using the noise maps and simulated point sources, each modelled as the point source kernel scaled by the source flux density. These simulated galaxies are injected at random positions drawn from a uniform distribution into the noise maps, where their flux density distribution is described by the best-fit Schechter function to the AzTEC/GOODS-S number counts (Section \ref{sec:nc}) truncated at $S_{1.1\rm{mm}}=0.1$\,mJy. These maps, referred to hereafter as ``fully simulated maps'', provide a realistic model of the mm-galaxy population in the GOODS-S field \textit{as observed by AzTEC} by properly including the effects from our data reduction methods. We use these fully simulated maps throughout this paper to investigate various properties of our map.

One might ask if the assumption of a spatially uniform population of sources in our simulations is justified. To date there are only weak constraints on the clustering properties of bright SMGs derived from incomplete and/or small-number samples, with significant disagreement among various surveys \citep[e.g.][]{blain04, greve04, scott06, weiss09}. Furthermore, bright sources like those probed in these surveys have low number density and so will not bias the map characteristics probed in our simulations. Rather, it is the faint, high-density, unexplored population of SMGs that can bias our estimates. Unfortunately, until better resolution for sub-mm/mm surveys is available (for example AzTEC on the LMT), we have no sufficient observational evidence to guide us. While these faint SMGs may have similar clustering properties as other high-redshift galaxy populations (e.g. LBGs, BzKs, EROs) it is likely that the projected 2-dimensional clustering strength of SMGs is very different from optically selected high-redshift galaxies, since sub-mm/mm surveys sample a much larger volume of space. In the absence of observational constraints on the clustering of faint SMGs we choose to model the source population with a uniform spatial distribution.

\subsection{False Detections}
\label{ssec:fdr}

\begin{figure}
\includegraphics{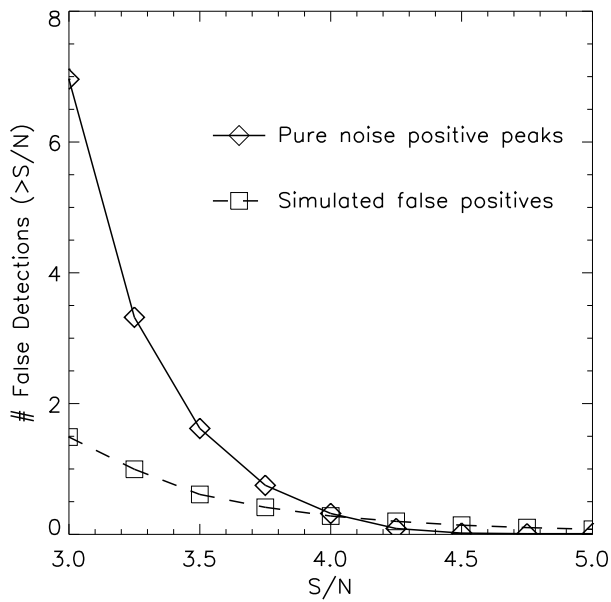}
\caption{The expected number of false detections in the AzTEC/GOODS-S map as a function of limiting $S/N$. The solid curve and diamonds show the number of false detections estimated from the number of peaks detected in pure noise realisations with significance $\ge S/N$. The dashed curve and squares indicate the expected number of false positives determined from fully simulated maps. The pure noise realisations provide only a conservative upper limit to the number of false detections expected.}
\label{fig:fdr}
\end{figure}

Given the modest $S/N$ of the source candidates, we expect some fraction of the AzTEC sources in GOODS-S to be spurious. One method commonly used to estimate the number of false positives is to run the source-finding algorithm on the \textit{negative} of the signal map. However, the PCA cleaning used for atmosphere removal leaves the signal map (and the point source kernel) with a mean of zero. Every real source that increases the number of positive pixels in the map will thus also increase the number of negative pixels, so the number of false detections measured from the negative of the real map will be overestimated. Instead, we estimate the number of false detections by identifying the number of ``sources'' extracted from the 100 pure noise realisations. The number of false detections expected as a function of limiting $S/N$ ratio is shown in Figure \ref{fig:fdr} (solid curve, diamonds). At $\ge3.5\sigma$, we expect $\sim2/40$ sources in our catalogue (5\%) to be spurious. Above $\ge4.25\sigma$, the number of AzTEC sources expected to be false positives is consistent with zero.

This estimate however provides only an upper limit to the number of spurious detections. In the real map, the negative bias in the pixel flux distribution from the addition of sources decreases the number of high-significance positive noise peaks in the map. This effect was first demonstrated for the AzTEC/GOODS-N survey \citep{perera08} and is even more pronounced for our confusion-limited GOODS-S map. To demonstrate this effect, we run the source-finding algorithm on 600 fully simulated maps: for each \textit{detected} source, we search within a $17\arcsec$ radius to identify the corresponding \textit{input} source, and detected sources that cannot be traced back to an input source are false positives. We choose a 17\arcsec~search radius to ensure recovery of an input source, since a source detected at $S/N\ge3.5$ has a probability $<0.1\%$ of being detected $>$17\arcsec~from its true position \citep{ivison07}. However given the high source density at faint flux densities, there is a non-negligible probability that an intrinsically faint source will be located within $17\arcsec$ of a \textit{random} position in the map, and so this calculation will underestimate the number of false positives. We account for the fraction of random associations as follows: 1) for each of the 600 fully simulated maps, we select $2,000$ random positions within the 50\% uniform coverage region, identify input sources located within $17\arcsec$ of these random positions, and from this determine the probability $P(>17\arcsec|S_i)$ that a source with intrinsic flux density $S_i$ will be located within $17\arcsec$ of a random position in the map; 2) for each detected source in the simulated maps that can be traced back to an input source with intrinsic flux density $S_i$, we generate a random number, $P_{\rm{test}}$, between 0 and 1 from a uniform distribution, and we consider the output-input source pair a false association (and hence the output source a false positive) if $P_{\rm{test}}<P(>17\arcsec|S_i)$.

The number of false positives estimated from these fully simulated maps as a function of limiting $S/N$ is shown in Figure \ref{fig:fdr} (dashed curve, squares). From this estimate we expect at most one of the 40 $\ge3.5\sigma$ sources in our catalogue to be spurious. At $\ge3.0\sigma$, the number of false positives estimated from these fully simulated maps is significantly lower than that estimated from pure noise realisations ($1-2$ versus $\sim7$), suggesting that we can comfortably extend our source catalogue to lower $S/N$ detections. However, these simulations do not include the effects of source clustering. If the mm-galaxy population is strongly clustered on small angular scales, the strength of the negative bias in the pixel flux distribution would vary from region to region due to the variance in the source density, and thus the number of positive noise peaks (i.e. false positives) would be lower (higher) in the more (less) densely populated regions of the map. Since the clustering properties of the mm-galaxy population (including intrinsically faint sources) are not well known, we prefer to quote the values determined from the pure noise realisations as a conservative estimate of the number of false detections expected in our catalogue.

\subsection{Completeness}
\label{ssec:completeness}

\begin{figure}
\includegraphics{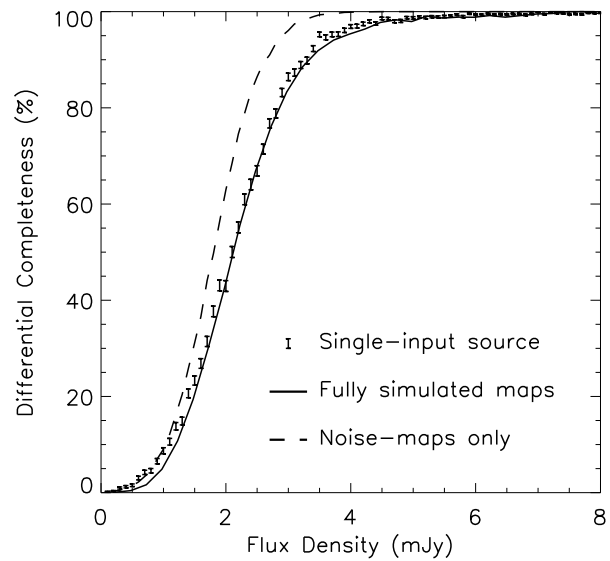}
\caption{The survey completeness for $S/N \ge3.5$ AzTEC sources in GOODS-S. The data-points and 68\% confidence binomial error bars show the completeness estimated by inserting sources of known flux density one at a time into the real signal map. The solid curve shows the completeness estimated by fully simulated maps. The dashed curve shows the completeness estimated by inserting sources of known flux density one at a time into pure noise realizations. This latter estimate does not account for the effects of confusion noise and hence results in an overestimate of the survey completeness for the range of flux densities shown here.}
\label{fig:completeness}
\end{figure}

The detection rate for a given source flux density is affected by both Gaussian random noise in the map and confusion noise from the underlying bed of faint sources. To account for both effects, we estimate the survey completeness by measuring the recovery rate of point sources with known flux densities inserted into the real signal map, as described in \citet{scott08}. For flux densities ranging from $0.1-8.0\,\rm{mJy}$, we input 2000 sources per flux density \textit{one at a time} into the GOODS-S map, each time randomly selecting the source position. Using the standard source-finding algorithm, an input source is considered recovered if it is detected in the map within $17\arcsec$ of its input position with a significance of $\ge3.5\sigma$. We exclude samples where the simulated source was input or extracted less than $17\arcsec$ from a real $\ge3.5\sigma$ source. The survey completeness is shown in Figure \ref{fig:completeness} (data-points with binomial error bars). The survey is 50\% complete at $2.1\,\rm{mJy}$, and 95\% complete at $3.5\,\rm{mJy}$.

The sea of faint sources below the detection threshold adds confusion noise to the AzTEC/GOODS-S map. This additional noise reduces the map's sensitivity to individual sources and its survey completeness over a range of flux densities. An accurate estimate of the completeness is essential for correcting the observed number counts for this field. In the standard Bayesian method for extracting number counts from AzTEC maps \citep{austermann09,austermann10}, survey completeness is estimated by injecting sources of known flux density one at a time into noise realisations and determining their recovery rate. While this method does not take confusion noise into account, this estimate was found to be consistent with that measured from the real signal map using the method described in the previous paragraph for several AzTEC surveys on the JCMT, where the angular resolution is better ($\rm{FWHM}=18\arcsec$ on the JCMT versus $30\arcsec$ on the ASTE) and the $1\sigma$ depths of the surveys are $\sim1.0\,\rm{mJy}$ \citep[e.g][]{perera08,austermann09,austermann10}, demonstrating that confusion noise was not significant for these observations. In contrast, we find that the completeness estimated from noise-only maps significantly over-predicts the survey completeness for our deeper, confusion-limited GOODS-S map (Figure~\ref{fig:completeness}, dashed curve).

To verify that this difference arises from confusion noise, we next estimate the survey completeness from $10,000$ fully simulated maps. For each $\ge3.5\sigma$ source detected in these simulated maps, we identify the brightest input source within 17" of the output source position. We bin all detected sources by their input flux densities, and the completeness is calculated by the ratio of the number of recovered sources to the total number of input sources per flux bin. The completeness estimated from these fully simulated maps is shown as the solid curve in Figure~\ref{fig:completeness}. This estimate agrees quite well with that from the single-input source simulations using the real GOODS-S map. The discrepancy between the two methods at $S_{1.1\rm{mm}}\lesssim1.5\,\rm{mJy}$ likely arises from small imperfections in the assumptions we use to identify input-output pairs. For example, the single-input source simulations may slightly overestimate the completeness at low flux densities due to cases when the input source is inserted close to (but $>17\arcsec$) a bright mm-galaxy in the real map. Still, despite the very different methods used for these two different completeness estimates, they differ by $\le4$\% at all flux densities.

\subsection{Positional Uncertainty}
\label{ssec:positional_uncertainty}

\begin{figure}
\includegraphics{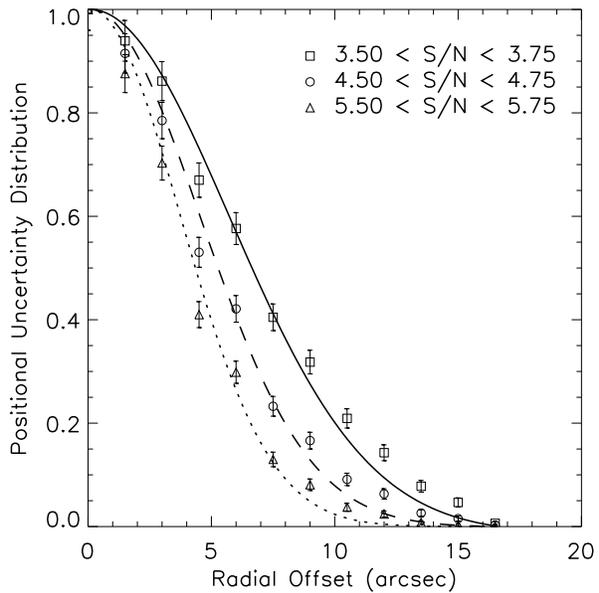}
\caption{The positional uncertainty distribution for AzTEC/GOODS-S source candidates. The data-points and error bars show the probability $P(>\theta; S/N)$ that an AzTEC source detected with a given $S/N$ ratio will be found outside a radial distance $\theta$ from its true location as determined from simulation (Section~\ref{ssec:positional_uncertainty}). The curves show the analytical expression derived in \citet{ivison07}: \textit{solid}, $3.5<S/N<3.75$; \textit{dashed}, $4.5<S/N<4.75$; and \textit{dotted}, $5.5<S/N<5.75$. The empirical and analytical distributions show broadly the same trend.}
\label{fig:pos_unc}
\end{figure}

The large beam combined with the low $S/N$ of the detections results in a large positional error on the locations of sub-mm/mm-detected sources due to the effects of random and confusion noise in the map. We use the simulations described in \ref{ssec:completeness}, where a single source of known flux density is inserted into the GOODS-S map one at a time, to determine the distribution of input to output source distances as a function of detected $S/N$. The probability $P(>\theta; S/N)$ that a source will be detected outside of a radial distance $\theta$ of its true position is shown in Figure~\ref{fig:pos_unc} for three sample $S/N$ bins. For comparison, the analytical solutions determined from \citet{ivison07} are shown as the solid ($3.5<S/N<3.75$), dashed ($4.5<S/N<4.75$), and dotted ($5.5<S/N<5.75$) curves, assuming that the FWHM of the AzTEC beam is $34\arcsec$ (i.e. the width of best-fit Gaussian to the \textit{filtered} point source kernel). The analytical and empirical distributions show broadly the same trend.

\subsection{Source Blending}
\label{ssec:confusion_sourceblend}

Given the depth and low angular resolution of the AzTEC/GOODS-S survey, some fraction of the $\ge3.5\sigma$ sources in the map are expected to be the combined signal from two or more individual galaxies blended together. To estimate the fraction of the sources in Table~\ref{table:sources} that are actually the blend of $\ge2$ individual galaxies, we take the $\ge3.5\sigma$ sources detected from 600 fully simulated maps and trace each one back to all input sources located within $17\arcsec$ of the output source position. The fraction of detected sources that cannot be traced back to any input source is 0.8\%: these represent the ``false positives" estimated from fully simulated maps as described in Section \ref{ssec:fdr}. Sources that map back to only one input source are considered ``single sources", while those that can be traced back to two or more input sources are considered ``blended sources".

With this simple definition, we would expect nearly all sources to be blended given the large beam and the high source density of SMGs. However, a very faint source nearby a relatively bright source (for example, a $0.2\,\rm{mJy}$ source located $10\arcsec$ from a $3.0\,\rm{mJy}$ source) contributes a negligible amount to the summed signal. We want to avoid counting cases like these -- where the brighter of the two sources completely dominates the total signal -- as a blended source. As a more practical definition, we only consider a pair of nearby sources to be a blend if the contribution from each source to the summed signal is comparable. Using the input source flux densities and relative separations for all simulated sources within $17\arcsec$ of an output source, we model the beam-smoothed \textit{noiseless} signal from the sum of these point sources and measure the peak flux density. If the fractional contribution to the summed flux density for an individual input source at the location of the peak is $\ge70$\%, we consider this a single source; otherwise, the detection is considered a blended source. With this definition, 25\% ($10/40$) of the AzTEC sources listed in Table~\ref{table:sources} are expected to be $\ge2$ individual galaxies blended by the large beam. This fraction reduces to 15\% if we define a detection as a single source when it contributes $\ge60$\% to the summed noiseless flux density. These results represent lower limits to the fraction of blended sources, since this fraction would be even higher if the SMG population is significantly clustered. We note that by design, this fraction does not depend on the limiting $S/N$ threshold (at least for $S/N\ge3.5$).

\subsection{Summary of Confusion Effects}
\label{ssec:sum_conf}

We summarise our primary findings on the effects of confusion on the map properties:

\begin{itemize}

\item When the signal from confused sources becomes comparable to that from random noise fluctuations, the number of pure-noise positive peaks in the map decreases. Thus the number of false positives as a function of limiting $S/N$ ratio is considerably lower than that estimated from peaks in noise-only realisations.

\item In terms of the survey completeness for a given catalogue, the confused signal acts as an additional noise component that reduces the map's sensitivity to individual galaxies over a wide range in intrinsic source flux density. It is important to include the effects of confusion when estimating the survey completeness for use in number counts calculations (Section~\ref{ssec:bayes}).

\item A potentially large fraction of sources detected in a confusion-limited map are actually multiple galaxies blended by the large beam. Care must be taken to understand the possible biases this could cause in number counts estimations (Section~\ref{ssec:bayes}).

\end{itemize}

\section{NUMBER COUNTS}
\label{sec:nc}

\subsection{Fluctuation Analysis}
\label{ssec:fluc_analysis}

\begin{figure*}
\includegraphics{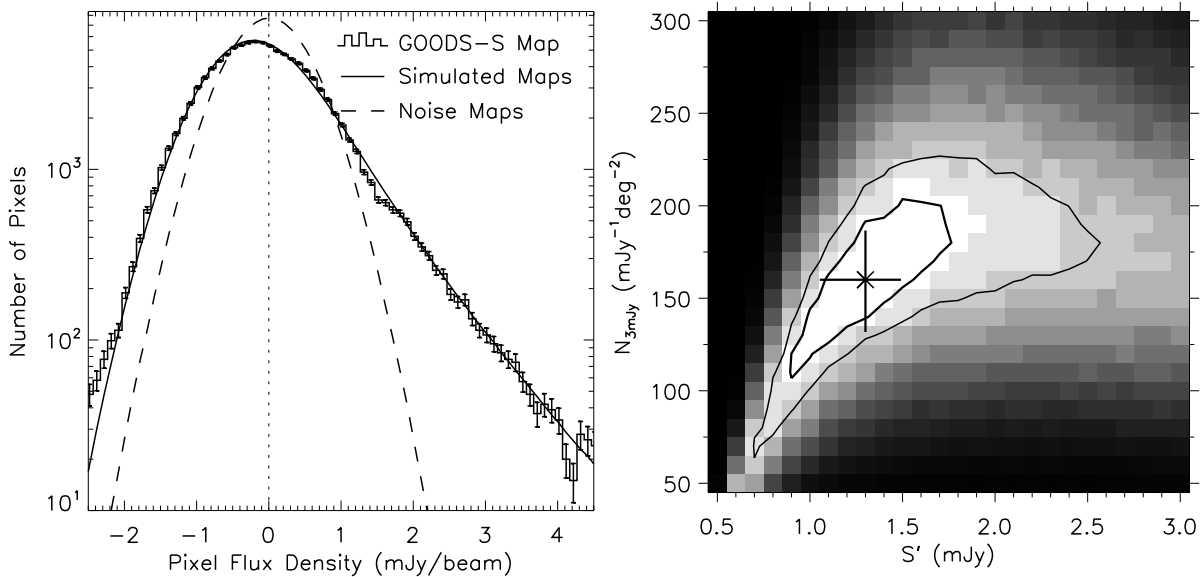}
\caption{\textit{Left}: The histogram of flux density values in the AzTEC/GOODS-S map. The dashed curve shows the distribution of flux values averaged over 100 noise realisations for this field, and is Gaussian distributed about zero. The solid curve shows the flux distribution averaged over fully simulated maps, populated according to the best-fit Schechter function model to the GOODS-S data. This demonstrates how a fluctuation analysis is used to determine the best model to the GOODS-S data. \textit{Right}: The grey-scale shows the likelihood values of $S^{\prime}$ and $N_{3\rm{mJy}}$ from the fluctuation analysis, and the cross indicates the best-fit parameters to the data. The inner and outer contours indicate the 68.3\% and 95.5\% confidence regions, respectively. The error bars represent marginalized 68.3\% confidence intervals on each parameter.}
\label{fig:flucanal}
\end{figure*}

Due to the exceptional depth reached by this survey, the mm-emission from faint SMGs has a striking effect on the flux density distribution in the map. As discussed already, the method used to remove low-frequency modes leaves the mean of the map and the point source kernel equal to zero, and every mm-source adds both positive and negative flux to the map. This is demonstrated in the left panel of Figure~\ref{fig:flucanal}, which shows the histogram of flux density values in the AzTEC/GOODS-S map (with Poisson error bars). The dashed curve shows the distribution of pixel fluxes averaged over 100 jackknifed noise realisations of this field, and is very Gaussian. The flux distribution in the real map on the other hand is skewed by the presence of mm-sources. While this effect makes the identification of individual galaxies challenging, we can use the distribution of flux values in the map to perform a fluctuation analysis, or more commonly referred to as a ``$P(d)$'' analysis, in order to determine the number counts distribution for this field. As this technique is independent of the level of confusion in the map (at the expense of being somewhat model dependent), we can use it on the AzTEC/GOODS-S map to provide strong constraints on the SMG number counts at faint flux densities ($S_{1.1\rm{mm}}<1\,\rm{mJy}$) well below the $1.1\,\rm{mm}$ confusion-limit.

The fluctuation analysis is carried out as follows: using a parametrised model of the number counts, we generate 100 simulated maps as described in Section~\ref{sec:confusion} and compare the flux density distribution averaged over these simulated maps to that of the real GOODS-S map. For this single model, we calculate the comparison metric:
\begin{equation}
\label{equ:likelihood}
-\ln({\cal L}) = \sum_i m_i - d_i + d_i \cdot \ln(d_i/m_i)
\end{equation}
where $m_i$ represents the average number of pixels in the $i^{\rm{th}}$ flux density bin from the model and $d_i$ represents the corresponding quantity for the GOODS-S map. This process is repeated over a grid in parameter space to find the minimum of the above metric, which occurs at the best-fit model. Minimising this metric is equivalent to finding the maximum likelihood for the case that all histogram bins follow {\it independent} Poisson distributions. Note that we do not attempt to model effects of source clustering on the pixel flux distributions, since the clustering properties of the SMG population are not well known. This method is similar in principle to the parametric frequentist approach used by \citet{perera08} to determine number counts for the AzTEC/GOODS-N survey; however, here we consider the full flux density histogram in order to extract information about the faint source population.

For this analysis we choose a Schechter function model given by:
\begin{equation}
\label{eqn:schechter}
{\mbox{d}N \over \mbox{d}S} = N_{3\rm{mJy}} \left({S \over 3\,\rm{mJy}}\right)^{\alpha+1} \mbox{exp}\left({-(S-3\,\mbox{mJy}) \over S^{\prime}}\right)
\end{equation}
where ${\mbox{d}N \over \mbox{d}S}$ is the differential number counts as a function of intrinsic 1.1\,mm flux density $S$, and $(S^{\prime},N_{3\rm{mJy}},\alpha)$ are the free parameters. While there are many forms of the Schechter function published in the literature, we prefer to fit to $N_{3\mathrm{mJy}}$ because it reduces the degeneracies between the parameters, and it has been used in the number counts analyses of several previous AzTEC surveys \citep{perera08, austermann09, austermann10}, making it straightforward to compare the results. We prefer a Schechter function model over that of a single power-law because it allows for a natural steepening of the counts at high flux densities, which has been confirmed by large-area surveys at sub-mm/mm wavelengths \citep{coppin06, austermann10} -- though we are unlikely to see this steepening in the GOODS-S number counts given the small survey area. Since our data cannot provide strong constraints on a 3-parameter fit, we fix $\alpha=-2$: a value that is consistent with estimates from previous surveys \citep[e.g.][]{coppin06, perera08, austermann09, austermann10}.

Since the Schechter function increases to infinity as $S$ goes to zero, we must assume some minimum flux density cutoff, $S_{\rm{min}}$, for the population. A practical minimum flux limit is imposed by the data itself: at the flux density corresponding to where the number density of sources is $\sim1$ per beam, adding fainter sources will not alter the flux density distribution in the map. Assuming the best-fit model to the AzTEC/SHADES data \citep[][which currently provides the tightest constraints on the blank-field $1.1\,\rm{mm}$ number counts]{austermann10}, $S_{\rm{min}}\sim0.1\,\rm{mJy}$. While it is not known whether the number density of sources for the SMG population turns over and starts to decrease somewhere below $1\,\rm{mJy}$, the counts at the faint end of the 850\,\micron~SCUBA galaxy population determined from lensing cluster surveys \citep[e.g.][]{cowie02, smail02, knudsen06, knudsen08} continue to rise out to $S_{850\mu\rm{m}}\sim0.2\,\rm{mJy}$, giving some reassurance that a $1.1\,\rm{mm}$ flux cutoff of $S_{\rm{min}}=0.1\,\rm{mJy}$ is reasonable. We use $S_{\rm{min}}=0.1\,\rm{mJy}$ in generating all simulated maps discussed in this paper; however, we have tested values ranging from $S_{\rm{min}}=0.05-0.3\,\rm{mJy}$ and have found that this does not affect the results from the fluctuation analysis.

Using $0.1\,\rm{mJy}$ bins for the flux histograms, we restrict the data-model comparison to bins with $\ge10$ pixels on average (flux densities between $-2.8$ and $5.5\,\rm{mJy}$). The resulting best-fit parameters are $(S^{\prime},N_{3\rm{mJy}}) = (1.30^{+0.19}_{-0.25}\,\rm{mJy},160^{+27}_{-28}\,\rm{mJy}^{-1}\rm{deg}^{-2})$. The flux distribution for this best-fit model is shown as the solid curve in the left panel of Figure~\ref{fig:flucanal}. The likelihood values for the $S^{\prime}-N_{3\rm{mJy}}$ parameter space are shown in the right panel of Figure~\ref{fig:flucanal}, with the best-fit parameters indicated by the cross. Due to the strong bin-to-bin correlations, it is not possible to determine the errors on the best-fit parameters analytically. Instead, we determine the errors statistically though simulation by generating 600 fully simulated maps populated assuming the best-fit Schechter function model to the real GOODS-S map (including Poisson deviations), and then performing the same fluctuation analysis on these simulated maps. The distribution of best-fit parameters from these simulated maps are used to determine the 68.3\% and 95.5\% confidence intervals (contours in Figure~\ref{fig:flucanal}). The errors given for $S^{\prime}$ and $N_{3\rm{mJy}}$ above (and shown as error bars in Figure~\ref{fig:flucanal}) represent the marginalized 68.3\% confidence intervals on each parameter.

\begin{figure}
\includegraphics{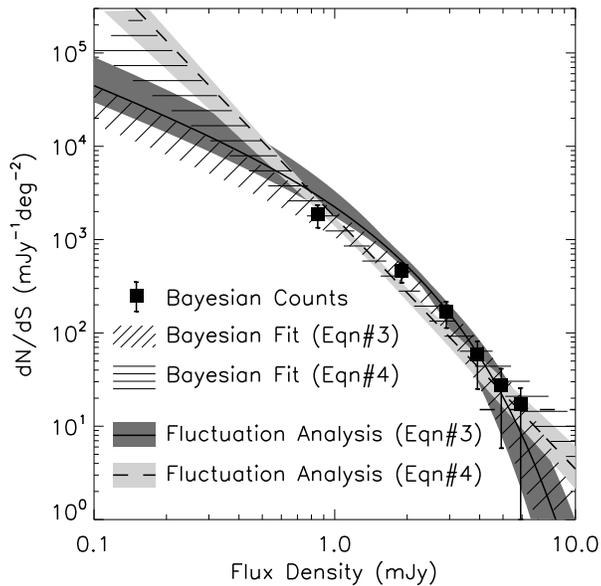}
\caption{The differential number counts for the AzTEC/GOODS-S field using various methods and models. The solid curve and dark shaded region indicate the best-fit model and 68.3\% confidence interval from the fluctuation analysis, assuming a Schechter function model. The dashed curve and light shaded region indicate the best-fit model and 68.3\% confidence interval from a fluctuation analysis assuming a single power-law model. The squares show the Bayesian-extracted number counts and their 68.3\% confidence intervals determined in Section~\ref{ssec:bayes}. The slanted (horizontal) hatching shows the 68.3\% confidence interval for a fit to the Bayesian-extracted counts, assuming a Schechter function (power-law) model for the population. The horizontal dashed line shows the survey limit, where the number counts will Poisson deviate to $0\,\rm{mJy}^{-1}\rm{deg}^{-2}$ 31.7\% of the time given the area of the GOODS-S survey.}
\label{fig:nc_methods}
\end{figure}

The ability to recover the input number counts determined from the fully simulated maps verifies the reliability of this fluctuation analysis method. The best-fit model to the actual GOODS-S data is listed in the first row of Table~\ref{table:ncfit}, and the differential number counts from this fluctuation analysis is shown in Figure~\ref{fig:nc_methods} as the thick solid curve (best-fit) and dark shaded region (68.3\% confidence interval). These results are also shown in Figures~\ref{fig:dnc} and \ref{fig:inc} for a comparison with other surveys.

\begin{table}
\caption{\label{table:ncfit}The best-fit parameters for models to the AzTEC/GOODS-S number counts. The method used is listed in the first column: ``$P(d)$'' = fluctuation analysis, and ``Bayes'' = Bayesian method. The errors on the best-fit parameters represent marginalised 68.3\% confidence intervals. When an error is not listed, the parameter was fixed to the given value.}
\begin{center}
\begin{tabular}{llccc}
\hline
\hline
       &                         & $S^{\prime}$           & $N_{3\rm{mJy}}$                & $\alpha$  \\ 
Method & Model                   & (mJy)                  & ($\rm{mJy}^{-1}\rm{deg}^{-2}$) &           \\
\hline
$P(d)$     & Eqn~\ref{eqn:schechter} & $1.30^{+0.19}_{-0.25}$ & $160^{+27}_{-28}$    & $-2.0$                  \\
Bayes      & Eqn~\ref{eqn:schechter} & $1.47^{+0.40}_{-0.25}$ & $131^{+30}_{-20}$    & $-2.0$                  \\
$P(d)$     & Eqn~\ref{eqn:powerlaw}  & $-$                    & $~90^{+20}_{-18}$    & $-3.70^{+0.18}_{-0.11}$ \\
Bayes      & Eqn~\ref{eqn:powerlaw}  & $-$                    & $107^{+30}_{-10}$    & $-3.35^{+0.25}_{-0.15}$ \\
\hline
\end{tabular}
\end{center}
\end{table}

While potentially providing tight constraints on the source counts of the SMG population, this fluctuation analysis relies strongly on the accuracy of our noise realisations and the point source kernel, and is sensitive to the assumed number counts model. To demonstrate this, we carry out a fluctuation analysis on the GOODS-S data assuming instead a single power-law model for the number counts, which is given by
\begin{equation}
\label{eqn:powerlaw}
{\mbox{d}N \over \mbox{d}S} = N_{3\rm{mJy}} \left({S \over 3\,\rm{mJy}}\right)^{\alpha+1}.
\end{equation}
We chose this functional form to make both parameters easily comparable to those from the Schechter function model, since $N_{3\rm{mJy}}$ gives the differential number counts at $3\,\rm{mJy}$, and $\alpha$ represents the power-law dependence at low flux densities. The best-fit parameters are listed in the third row of Table~\ref{table:ncfit}, and the best-fit model and 68.3\% confidence interval are shown as the dashed curve and light shaded region in Figures~\ref{fig:nc_methods}, \ref{fig:dnc}, and \ref{fig:inc}. The best-fit values of $N_{3\rm{mJy}}$ for the two different models are only marginally consistent, and the best-fit power-law index ($\alpha+1=-2.7$) is significantly steeper than that assumed for the Schechter function model ($\alpha+1=-1.0$). Comparing the two models in Figure~\ref{fig:nc_methods}, they disagree significantly at $S\sim2\,\rm{mJy}$, where supposedly the tightest constraints can be placed on the number counts. One must thus exercise caution when interpreting results from this fluctuation analysis, as it inherently assumes that the model provides a good representation of the source counts.

\subsection{Bayesian Estimation}
\label{ssec:bayes}

We can minimise the dependence of the analysis on our underlying model of the number counts by utilising a modified form of the semi-Bayesian method introduced by \citet{coppin05,coppin06}. This semi-Bayesian method is now widely used to extract the number counts from sub-mm/mm surveys because of its ability to handle various survey biases, and it has been extensively tested and validated using previous AzTEC data-sets \citep{perera08,austermann09,austermann10}. Since this method is described in detail in the aforementioned papers we only briefly summarise the steps here.

The raw source counts from sub-mm/mm surveys suffer from three main biases: the ``flux boosting'' effect described in Section~\ref{ssec:sources}, contamination from false positives, and incompleteness. To account for the first two effects, we generate posterior flux distributions (PFDs), $p(S_i|S_m,\sigma_m)$ (where $S_i$ is the intrinsic flux density of the source, $S_m$ is the measured flux density, and $\sigma_m$ is the error on the measured flux density), for each source candidate assuming some prior model for the SMG number counts. These PFDs are then randomly sampled (with replacement) to determine \textit{intrinsic} flux densities for the sources, and these fluxes are binned to calculate differential and cumulative number counts. This process is repeated $20,000$ times to adequately sample the number counts distribution. We also include sample variance by Poisson deviating the number of sources sampled in each of the $20,000$ iterations. Since the PFD for each source candidate includes a non-negligible probability that the intrinsic flux density is $S_i<0\,\rm{mJy}$, this procedure inherently accounts for false positives.

To extract source counts from the AzTEC/GOODS-S map, we use the best-fit Schechter function model determined from the fluctuation analysis as the prior distribution to generate PFDs for the source candidates. We sample all source candidates where the probability of the source having a flux density less than zero is $p(S_i<0|S_m,\sigma_m)\le0.05$, which corresponds to $S/N\gtrsim2.8$ (actually depends on both $S_m$ and $\sigma_m$) for a total of 54 source candidates. \citet{austermann10} tested various limiting thresholds for $p(S_i<0|S_m,\sigma_m)$ using several AzTEC data-sets and found that any variations in the resulting number counts are much smaller than the formal 68.3\% uncertainties. They found that for accurate PFDs, the $p(S_i<0|S_m,\sigma_m)$ limiting threshold does not greatly affect the resulting number counts (provided source confusion is not an issue), and that using a higher limiting threshold supplies more information (due to increased survey completeness at faint flux densities) without introducing significant biases. We have verified that the PFDs for GOODS-S source candidates with $S/N\ge2.8$ are accurate to better than $1$\% at $S_i\ge0.5\,\rm{mJy}$ following the simulations described in \citet{austermann09,austermann10}, further justifying the use of a $p(S_i<0|S_m,\sigma_m)\le0.05$ limiting threshold in this analysis. We note that while \citet{austermann10} use a limiting threshold of $p(S_i<0|S_m,\sigma_m)\le0.2$, we choose a more conservative limit to avoid potential systematics arising from confusion effects, which were negligible for \citet{austermann10}.

The raw number counts must also be corrected for incompleteness. In previous implementations of the semi-Bayesian method on AzTEC data-sets \citep{perera08,austermann09,austermann10}, survey completeness was estimated by the recovery rate of synthetic point sources with known intrinsic flux densities inserted into pure noise realisations one at a time. However, we have shown in Section~\ref{ssec:completeness} that this method overestimates the completeness for the AzTEC/GOODS-S field, since (unlike the AzTEC/JCMT surveys) confusion noise is significant for our data. Using a pure noise completeness estimate would consequently underestimate the number counts in this field. We instead estimate the survey completeness through fully simulated maps as described in Section~\ref{ssec:completeness}, where the simulated maps are populated according to the number counts distribution given by the assumed \textit{prior}. To be consistent with our catalogue selection from the real GOODS-S map, an input source in these simulations is considered to be recovered only if its corresponding output source passes the limiting threshold test of $p(S_i<0|S_m,\sigma_m)\le0.05$. Since we are using an ``ideal'' prior determined from the fluctuation analysis of this field, we are confident that this provides a good completeness estimate for the correction of the raw number counts.

\begin{figure*}
\includegraphics{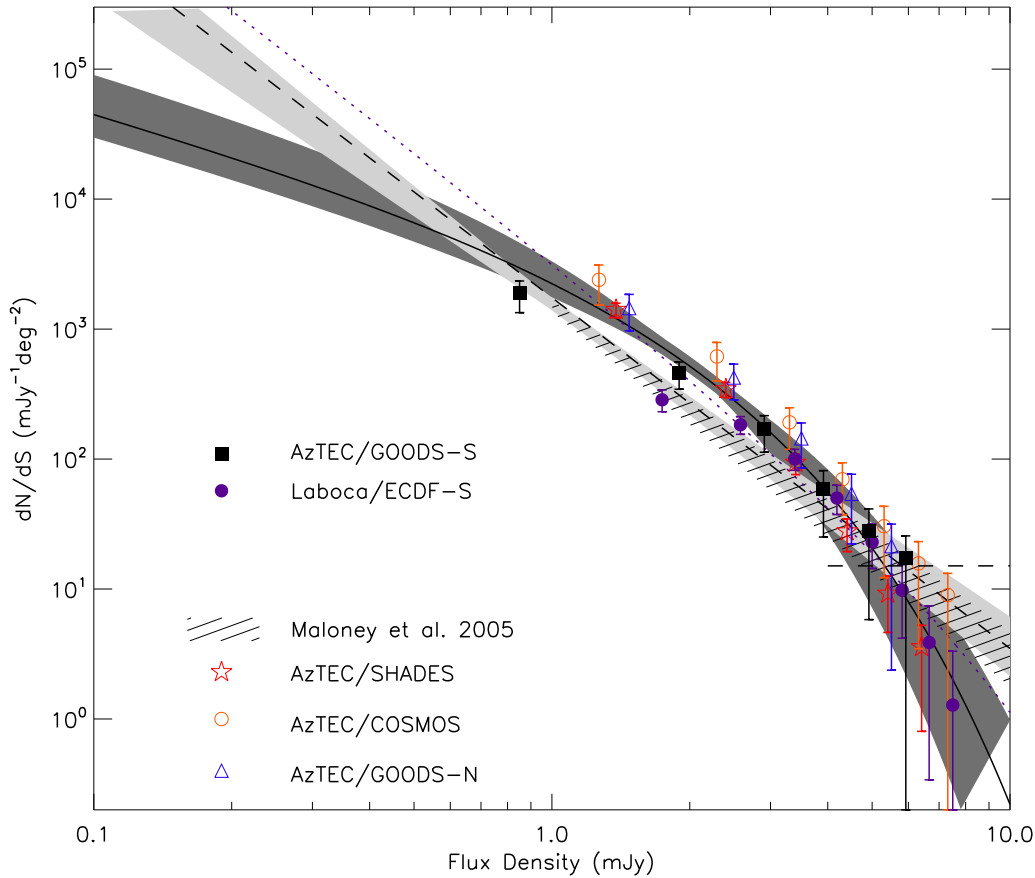}
\caption{The differential number counts from the AzTEC/GOODS-S survey (\textit{black squares}), compared with those determined from AzTEC surveys of other fields, including: GOODS-N \citep[][]{perera08}, COSMOS \citep[][]{austermann09}, and SHADES \citep[][]{austermann10}. The error bars represent 68.3\% confidence intervals on Bayesian-extracted counts. The solid (dashed) curve and dark (light) shaded region indicate the best-fit Schechter function (power-law) model and 68.3\% confidence region from a fluctuation analysis of GOODS-S (Section~\ref{ssec:fluc_analysis}). The horizontal dashed line shows the survey limit, where the number counts will Poisson deviate to $0\,\rm{mJy}^{-1}\rm{deg}^{-2}$ 31.7\% of the time given the area of the GOODS-S survey. The 95\% confidence interval from a fluctuation analysis of the 1.1\,mm Bolocam Lockman Hole survey \citep{maloney05} assuming a single power-law model is shown by the hatched region. The filled circles indicate the differential number counts determined from a Bayesian analysis of the 870\,\micron~LABOCA survey of the ECDF-S, scaled to 1.1\,mm assuming a flux ratio of $S_{870\mu\rm{m}}/S_{1.1\rm{mm}}=2.0$. The dotted curve shows the best-fit Schechter function model to the LABOCA/ECDF-S data from a $P(d)$ analysis scaled to 1.1\,mm (confidence region not available). See the on-line journal for a colour version of this Figure.}
\label{fig:dnc}
\end{figure*}

\begin{figure*}
\includegraphics{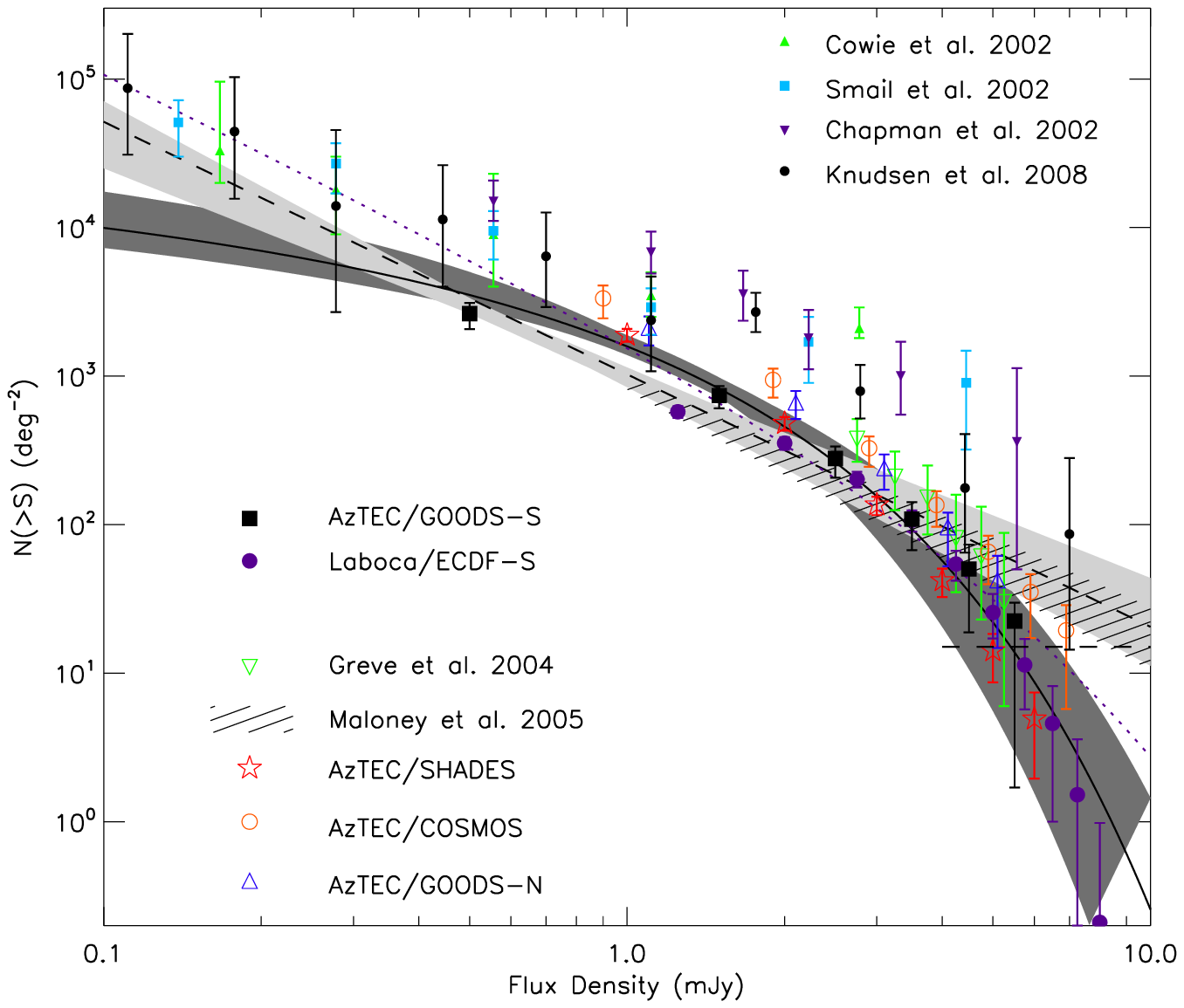}
\caption{The cumulative number counts from the AzTEC/GOODS-S survey and AzTEC surveys of other fields. See the caption of Figure~\ref{fig:dnc} for a full description of the symbols. The 1.2\,mm number counts from \citet{greve04} are shown as empty upside-down triangles with 95\% confidence error bars. For comparison, the number counts from SCUBA 850\,\micron~lensing cluster surveys are shown with smaller symbols. All 850\,\micron~counts have been scaled to $1.1\,\rm{mm}$ assuming a simple flux ratio scaling of $S_{850\mu\rm{m}}/S_{1.1\rm{mm}}=1.8$. The number counts from \citet{cowie02}, \citet{smail02}, \citet{chapman02}, and \citet{greve04} have not been corrected for flux boosting are therefore are likely biased to the right. See the on-line journal for a colour version of this Figure.}
\label{fig:inc}
\end{figure*}

The $1.1\,\rm{mm}$ differential number counts for the GOODS-S field determined from the Bayesian method are shown in Figures~\ref{fig:nc_methods} and \ref{fig:dnc} (filled squares), and the cumulative number counts are shown in Figure~\ref{fig:inc}. The counts are also listed in Table~\ref{table:bayesnc}. The number counts are calculated from the mean number of sources in each flux bin (with bin-size $=1\,\rm{mJy}$) over the $20,000$ iterations, and the errors represent the 68.3\% confidence intervals calculated from the distribution in the counts across those iterations. For the differential number counts, the flux densities in Table~\ref{table:bayesnc} are the effective bin centers weighted by the assumed prior. The number counts from this method are highly correlated since they are estimated by averaging over many realizations of the number counts bootstrapped off the \textit{same} source catalogue. The covariance matricies for the differential and cumulative number counts are listed is Tables~\ref{table:diffcov} and \ref{table:cumcov}, respectively.

\begin{table}
\caption{\label{table:bayesnc}The differential and cumulative number counts for the AzTEC/GOODS-S field, calculated using the Bayesian method described in Section~\ref{ssec:bayes}. The errors indicate 68.3\% confidence intervals.}
\begin{center}
\begin{tabular}{cr@{}lcr@{}l}
\hline
\hline
Flux Density & \multicolumn{2}{c}{$\mbox{d}N/\mbox{d}S$}          & Flux Density & \multicolumn{2}{c}{$N(>S)$}            \\ 
(mJy)        & \multicolumn{2}{c}{($\rm{mJy}^{-1}\rm{deg}^{-2}$)} & (mJy)        & \multicolumn{2}{c}{($\rm{deg}^{-2}$)}  \\ 
\hline
0.85         & $1892$ & $^{+453}_{-554}$   & 0.50          & $2631$ & $^{+478}_{-562}$ \\
1.90         & $ 461$ & $^{+ 97}_{-116}$   & 1.50          & $ 739$ & $^{+117}_{-132}$ \\
2.91         & $ 170$ & $^{+ 46}_{- 57}$   & 2.50          & $ 279$ & $^{+ 57}_{- 72}$ \\
3.92         & $  58$ & $^{+ 23}_{- 33}$   & 3.50          & $ 109$ & $^{+ 33}_{- 41}$ \\
4.92         & $  28$ & $^{+ 14}_{- 22}$   & 4.50          & $  50$ & $^{+ 23}_{- 31}$ \\
5.92         & $  17$ & $^{+  8}_{- 17}$   & 5.50          & $  22$ & $^{+  7}_{- 21}$ \\
\hline
\end{tabular}
\end{center}
\end{table}

\begin{table}
\caption{\label{table:diffcov}The covariance matrix for the differential number counts for the AzTEC/GOODS-S field. The units are in $\rm{mJy}^{-2}\rm{deg}^{-4}$.}
\begin{center}
\begin{tabular}{c|rrrrrr}
\hline
\hline
Flux Density & \multicolumn{6}{r}{} \\
(mJy)        & \multicolumn{0}{c}{0.85}   & \multicolumn{0}{c}{1.90} & \multicolumn{0}{c}{2.91} & \multicolumn{0}{c}{3.92} & \multicolumn{0}{c}{4.92} & \multicolumn{0}{c}{5.92} \\
\hline
0.85         & 254800 & 37340 & 3559 &  82 & -79 &  11 \\
1.90         &        & 11420 & 2835 & 345 &   8 &   8 \\
2.91         &        &       & 2623 & 848 &  73 &   5 \\
3.92         &        &       &      & 808 & 353 &  27 \\
4.92         &        &       &      &     & 366 & 127 \\
5.92         &        &       &      &     &     & 230 \\
\hline
\end{tabular}
\end{center}
\end{table}

\begin{table}
\caption{\label{table:cumcov}The covariance matrix for the cumulative number counts for the AzTEC/GOODS-S field. The units are in $\rm{deg}^{-4}$.}
\begin{center}
\begin{tabular}{c|rrrrrr}
\hline
\hline
Flux Density & \multicolumn{6}{r}{} \\
(mJy)        & \multicolumn{0}{c}{0.50} & \multicolumn{0}{c}{1.50} & \multicolumn{0}{c}{2.50} & \multicolumn{0}{c}{3.50} & \multicolumn{0}{c}{4.50} & \multicolumn{0}{c}{5.50} \\
\hline
0.50         & 271300 & 16060 & 4350 & 1478 & 659 & 307 \\
1.50         &        & 15580 & 4144 & 1470 & 650 & 300 \\
2.50         &        &       & 4123 & 1486 & 667 & 293 \\
3.50         &        &       &      & 1471 & 667 & 297 \\
4.50         &        &       &      &      & 670 & 301 \\
5.50         &        &       &      &      &     & 297 \\
\hline
\end{tabular}
\end{center}
\end{table}

The Bayesian-extracted number counts agree within the $1\sigma$ errors with the best-fit Schechter function model from the fluctuation analysis. However, the number counts in the two lowest flux density bins ($0.5-1.5\,\rm{mJy}$ and $1.5-2.5\,\rm{mJy}$) are low compared to the best-fit curve from the fluctuation analysis, while the number counts in the highest bin ($5.5-6.5\,\rm{mJy}$) are high. This may arise from a systematic bias in the Bayesian-extraction method due to source blending, since this technique does not account for the possibility that an individually detected source is the summed flux density of two or more galaxies. This would indeed result in the apparent bias seen here, as the number counts would be overestimated at high flux densities and underestimated at low flux densities. We have checked for this potential bias in the Bayesian method by running this analysis on source catalogues extracted from 600 fully simulated maps populated according to the best-fit model from the fluctuation analysis. Since in this case we know the exact form of the source counts input into each map, we can search for such effects in the output number counts. We find that the output differential number counts for these simulated maps do indicate that this bias due to source blending is present in the data: for the $5.5-6.5\,\rm{mJy}$ flux density bin, the extracted number counts are higher than the input number counts for 65\% of the simulated maps, while for the $0.5-1.5\,\rm{mJy}$ bin, the extracted counts are lower than the input counts for 60\% of the simulated maps. However, at all flux densities, the extracted number counts agree with the input number counts within their $1\sigma$ ($2\sigma$) errors at least 86\% (96\%) of the time, so this bias is small compared to the formal Poisson errors.

To verify that the Bayesian-extracted number counts are insensitive to the assumed prior, we reran this procedure on the AzTEC/GOODS-S map using a prior distribution that is consistent with a fit to the number counts from the AzTEC/SHADES survey: $(S^{\prime},N_{3\rm{mJy}},\alpha)=(1.11\,\rm{mJy},153\,\rm{mJy}^{-1}\rm{deg}^{-2},-2.0)$. We find that the extracted numbers counts using these two different priors agree within 4\% at flux densities $\ge1.5\,\rm{mJy}$. For the lowest flux bin at $0.5-1.5\,\rm{mJy}$, the results agree within 19\%, i.e. well within the formal $1\sigma$ error. This demonstrates that the results from this technique are robust given a reasonable assumption for the prior number counts distribution. For this reason, we can fit these number counts to various models.

For a given model, we fit to each of the $20,000$ bootstrap iterations separately as the flux density bins for a given iteration are uncorrelated \citep[see][]{austermann10}, and we use the distribution of best-fit parameters to determine the most likely values and their confidence intervals. The results of a fit to the GOODS-S differential number counts assuming a Schechter function model (Equation~\ref{eqn:schechter}) and a power-law model (Equation~\ref{eqn:powerlaw}) are given in rows 2 and 4 of Table~\ref{table:ncfit} and are shown in Figure~\ref{fig:nc_methods} as the hatched regions, which indicate the 68.3\% confidence intervals. The fits to the Bayesian-extracted counts are in good agreement with the results from the fluctuation analysis for a given model. To compare the two models, we compute the $\chi^2$ metric given by
\begin{equation}
\label{equ:chi2}
\chi^2 = (\textbf{d} - \textbf{m})~\textbf{w}~(\textbf{d} - \textbf{m})^{\rm{T}}
\end{equation}
where $\textbf{d}$ is the row-vector containing the differential number counts from the Bayesian method, $\textbf{w}$ is the corresponding weight matrix calculated from the inverse of the covariance matrix, and $\textbf{m}$ is the model number counts evaluated at the same flux density bins as $\textbf{d}$. For the best-fit Schechter function to the Bayesian-extracted number counts, $\chi^2=0.84$. This model provides a better fit to the data than the single power-law model, for which $\chi^2=5.4$. Note that since the errors are not Gaussian-distributed, this metric is not expected to be follow the $\chi^2$-distribution; we use this metric simply to compare the relative goodness-of-fit for these two models.

\subsection{Comparison with Results from SCUBA Lensing Cluster Surveys}
\label{ssec:lenscompare}

The AzTEC/GOODS-S survey currently provides the only constraints on the $1.1\,\rm{mm}$ number counts at $S_{1.1\rm{mm}}<1\,\rm{mJy}$. For comparison, none of the existing blank-field SCUBA surveys constrain the 850\,\micron~counts to comparatively faint flux densities ($S_{850\mu\rm{m}}\lesssim2\,\rm{mJy}$): the deepest \citep[e.g.][]{hughes98, eales00} are too small in area to provide statistically significant samples of faint SMGs, while the large-area surveys \citep[e.g.][]{borys03,coppin06} do not reach sufficient depths to probe such faint sources. The most significant constraints on the $S_{850\mu\rm{m}}\lesssim2\,\rm{mJy}$ number counts come from SCUBA lensing cluster surveys, where massive foreground clusters are used to probe faint background SMGs via gravitational lensing \citep[e.g.][]{smail02, cowie02, chapman02, knudsen06, knudsen08}. This is a powerful technique for detecting intrinsically faint high-redshift SMGs and for estimating their source counts at low flux densities, as the foreground clusters magnify both the flux of the background sources (by factors of typically $2-3$) and the area in the source plane, effectively decreasing the survey confusion-limit that hinders the sensitivity of blank-field observations. With new constraints on the $S_{1.1\rm{mm}}=0.5\,\rm{mJy}$ number counts from the AzTEC/GOODS-S survey, the faint end of the number counts determined from lensing cluster surveys can be compared to the results from a blank-field. Given the small sample size from lensing cluster surveys, this comparison is limited to the cumulative number counts.

The number counts from four separate SCUBA lensing cluster surveys are shown in Figure~\ref{fig:inc}.  To compare these with our $1.1\,\rm{mm}$ results, we must account for the difference in observing wavelengths. We assume here that SCUBA and AzTEC are sampling the same underlying population of sub-mm/mm-bright galaxies, and that the difference in the observed number counts can be described by a simple flux scaling: $R=S_{850\mu\rm{m}}/S_{1.1\rm{mm}}$. In principle, surveys at $1\,\rm{mm}$ may preferentially select sources at higher redshifts -- or sources with  colder dust temperatures -- than surveys at 850\,\micron. While there is some evidence that $1\,\rm{mm}$ surveys select on average higher redshift galaxies than 850\,\micron~surveys \citep{eales03, younger07, younger09, greve08}, other studies suggest that there is no significant difference between the two populations \citep{greve04, ivison05, bertoldi07}. 

A recent source-to-source comparison of SCUBA and AzTEC sources in the GOODS-N field \citep{chapin09} reveals that while the redshift distribution of $1.1\,\rm{mm}$ sources in that field peaks at a higher redshift than that of  850\,\micron~sources ($z=2.7$ versus $z=2.0$), the population is consistent with an average flux scaling with $R=1.8$. Alternatively, the 850\,\micron~number counts determined from the SCUBA/SHADES survey \citep{coppin06} and the $1.1\,\rm{mm}$ number counts from the GOODS-N field \citep{perera08} are consistent assuming a flux scaling of $R=2.1\pm0.2$. These estimates are equivalent to the expected flux ratio of a $z=2.5$ galaxy whose SED can be modelled as a single component modified black-body with $T_{d}=30\,\rm{K}$ and emissivity index $\beta=1.5$. As this model is consistent with the expected SEDs of local galaxies observed with the Infrared Astronomical Satellite (\textit{IRAS}) and SCUBA \citep{dunne00,dunne01} as well as the measured SEDs of several SMGs \citep{chapman05,kovacs06,pope06,coppin08}, we start by adopting a simple scaling factor of $R=1.8$ for the purposes of comparing the number counts from the SCUBA lensing cluster surveys to the AzTEC/GOODS-S number counts.

The scaled number counts from the SCUBA lensing cluster surveys agree only marginally with the number counts from the AzTEC/GOODS-S survey and are systematically higher at all flux densities (see Figure~\ref{fig:inc}). In fact, the lensing cluster survey number counts are systematically higher than those from {\it all} AzTEC blank-field surveys.  We first examine the possibility that the flux scaling factor of $R=1.8$ derived from the GOODS-N field by \citet{chapin09} is not universal by deriving the value of $R$ that minimises the residuals between the SCUBA lensing cluster number counts and the best-fit model to the 1.1\,mm number counts from the AzTEC/GOODS-S field.  The best-fit values and 68.3\% confidence errors for $R$ determined from each of the four lensing cluster data-sets shown in Figure~\ref{fig:inc} are given in Table~\ref{table:fluxratio}.  We find $R\ge3.3$ for all lensing cluster surveys considered here, regardless of which model (Schechter function or power-law) we consider.  Such high values of $R$ are inconsistent with observed values for individual bright SMGs (when both 850\,\micron~and $1.1\,\rm{mm}$ measurements are available) as well as predicted values assuming SEDs and redshifts typical of this population. We note however that a high value of $R=2.5\pm0.1$ has been estimated from a similar scaling of the 850\,\micron~and $1.1\,\rm{mm}$ number counts from the SCUBA and AzTEC/SHADES surveys \citep{austermann10}.  If $1.1\,\rm{mm}$ surveys are tracing a population of galaxies with either higher redshifts or colder dust temperatures than those detected at 850\,\micron, we would expect this flux ratio to be \textit{lower}, since these two bands probe closer to the dust peak. The high value of $R$ measured from scaling the SHADES number counts is likely due to systematics caused by differences in the number counts analyses or calibration, and we consider this an upper limit to the true flux ratio. Estimates of $R$ from scaling the 850\,\micron~lensing cluster number counts to the best-fit model to the AzTEC/GOODS-S data-set represent $\gtrsim2\sigma$ deviations from this upper limit of $R=2.5$ from the SHADES results, implying that even greater systematics exist in the lensing cluster number counts. Evidence for significant systematics in the derivation of the lensing cluster number counts is also given by the fact that the cumulative number counts at $S_{850\mu\rm{m}}>2\,\rm{mJy}$ from lensing cluster surveys are also higher than those from large-area blank-field SCUBA surveys \citep[e.g.][]{borys03,webb03,scott06,coppin06}.

\begin{table}
\caption{\label{table:fluxratio}The 850\,\micron~to $1.1\,\rm{mm}$ flux ratio estimated by scaling the number counts from SCUBA lensing cluster surveys to the best-fit model to the AzTEC/GOODS-S number counts. The columns are: 1) reference for the lensing cluster survey (last row is the results for all four surveys combined); 2) best-fit flux ratio $R=S_{850\mu\rm{m}}/S_{1.1\rm{mm}}$ and 68.3\% confidence iterval assuming the Schechter function model for the GOODS-S number counts (Equation~\ref{eqn:schechter}); and 3) best-fit flux ratio and 68.3\% confidence interval assuming the power-law model for the GOODS-S number counts (Equation~\ref{eqn:powerlaw}).}
\begin{center}
\begin{tabular}{l|cc}
\hline
\hline
Reference                                     & $R$ (Schechter)     & $R$ (Power-law)     \\
\hline
\citet{cowie02}                               & $5.6^{+0.9}_{-0.9}$ & $3.5^{+0.4}_{-0.5}$ \\
\citet{smail02}                               & $4.4^{+0.9}_{-0.8}$ & $3.2^{+0.3}_{-0.4}$ \\
\citet{chapman02}                             & $5.2^{+0.8}_{-0.7}$ & $5.5^{+0.5}_{-0.5}$ \\
\citet{knudsen08}                             & $3.3^{+0.4}_{-0.4}$ & $3.3^{+0.4}_{-0.5}$ \\
~~~($S_{850\micron}\le2$\,mJy) & $3.5^{+1.8}_{-1.4}$ & $2.8^{+0.6}_{-0.7}$ \\
All                                           & $4.0^{+0.3}_{-0.3}$ & $3.5^{+0.2}_{-0.3}$ \\
\hline
\end{tabular}
\end{center}
\end{table}

There are several issues in the number counts extraction from lensing cluster surveys which may systematically bias the number counts high. Errors in the detailed mass models of the lensing clusters are an obvious candidate for the observed bias; however, any errors in estimating magnification factors for source flux densities are compensated by equivalent errors in the source plane area \citep{knudsen06, knudsen08}. Contamination by foreground cluster members is another concern, but these have been identified and excluded in these lensing cluster analyses. All of the lensing cluster surveys considered here include sources detected with low significance ($S/N\ge3.0$), and thus could include a significant number of false positives which would bias the number counts high. However, we believe that the dominate systematic that causes the discrepancies between our number counts and those derived from these lensing cluster surveys is due to the fact that none of these analyses, excepting that of \citet{knudsen08}, account for the flux boosting effect described in Section~\ref{ssec:sources}. \citet{knudsen08} apply a correction for flux boosting only for the sources detected in their deepest maps ($S_{850\mu\rm{m}}\lesssim2\,\rm{mJy}$). If we fit for $R$ using only the \citet{knudsen08} number counts at $S_{850\micron}\le2$\,mJy (row 5 in Table~\ref{table:fluxratio}), the best-fit value agrees with our conservative upper limit of $R=2.5$ within the $1\sigma$ errors. However, this effect arises from the steep number counts distribution for SMGs: it is more likely that a source detected in a flux-limited survey is an intrinsically faint source (numerous) boosted high by noise than an intrinsically bright source (rare). Flux boosting is thus prominent in \textit{all} maps, and we believe that the \citet{knudsen08} number counts have not been fully corrected for this effect. Still, we note that if SMGs cluster strongly on small angular scales, the apparent bias in lensing cluster surveys may simply be the result of cosmic variance.  The total numbers of lensing clusters analysed is still small, and the errors on the cumulative counts for the lensing cluster surveys are dominated by Poisson noise due to the limited survey areas ($\le40\,\rm{arcmin}^2$ in the source plane) and small sample sizes. 

\subsection{Comparison with Other $1.1\,\rm{mm}$ Surveys}
\label{ssec:aztecsurveys}

The differential and cumulative number counts from other $1.1\,\rm{mm}$ blank-field surveys are shown in Figures~\ref{fig:dnc} and \ref{fig:inc} for comparison. These include the AzTEC surveys of GOODS-N \citep{perera08}, COSMOS \citep{scott08, austermann09}, and SHADES \citep{austermann10} taken on the JCMT. The SHADES data-set consists of two separate regions of sky -- the Lockman Hole-East and the Subaru/XMM-Newton Deep Field -- and covers a total area of $0.67\,\rm{deg}^2$, making it the largest blank-field survey at $1.1\,\rm{mm}$ published to date. The 95\% confidence interval from a fluctuation analysis of the 1.1\,mm Bolocam Lockman Hole survey \citep[$1\sigma=1.4$\,mJy;][]{laurent05,maloney05}, where a single power-law model is assumed, is also shown. In Figure~\ref{fig:inc}, we show the 1.2\,mm cumulative number counts derived from the MAMBO surveys of the Lockman Hole and ELAIS N2 fields \citep{greve04}.

The best-fit single power-law model for the 1.1\,mm number counts determined from our fluctuation analysis of the AzTEC/GOODS-S map is in good agreement with that determined from the fluctuation analysis of the Bolocam Lockman Hole survey. However, given that the AzTEC/GOODS-S survey is $2-3$ times deeper than the Bolocam LH survey with roughly the same area, we are sceptical of the extremely tight confidence intervals reported by \citet{maloney05}. The single power-law model has been shown to poorly describe the SMG number counts over a wide range in flux density \citep[e.g.][]{scott06, coppin06, austermann10}, with the best-fit slope likely a compromise between the steeper slope at high flux densities and the shallower slope at low fluxes. The resulting model inflexibility is a likely contributor to the underrepresented errors. While the 1.2\,mm number counts from \citet{greve04} are broadly consistent with the number counts determined from several AzTEC surveys, we note that this analysis does not fully account for the substantial flux boosting effect on the measured source flux densities.

The general agreement among the $1.1\,\rm{mm}$ blank-field surveys taken with AzTEC is quite good. The identical, well-tested methods used to reduce and analyse these data-sets minimises any systematic differences among the results, so this agreement is naturally expected. For the same reason, any subtle differences seen among these surveys is significant. The number counts from the GOODS-S field are consistent with those from the SHADES survey, which currently provides the tightest constraints on the blank-field $1.1\,\rm{mm}$ number counts over this flux density range. In contrast, the GOODS-N and COSMOS fields appear somewhat over-dense by as much as $\sim50$\%; we see a significant amount of variation in the number counts on $\sim100\,\rm{arcmin}^2$ scales. Since galaxies are organized within structures with significant power on comoving scales $\lesssim100\,\rm{Mpc}$ \citep[e.g. ][]{springel05}, variations in the observed number counts is expected for surveys covering $\lesssim1\,\rm{deg}^2$.\footnote{$10\arcmin=5.1\,\rm{Mpc}$ at $z=2$ in the concordance model of cosmology.} A quantitative study of the degree of field-to-field variations observed among all blank-field surveys taken with AzTEC on the JCMT and the ASTE will be presented in a future paper (Scott et al. in prep.).

\section{Comparison with the LABOCA Survey of the ECDF-S at 870\,\micron}
\label{sec:laboca}

A $0.35$\,deg$^2$ region that covers the Extended Chandra Deep Field South (ECDF-S) has been imaged at 870\,\micron~with the LABOCA camera on the APEX telescope \citep{weiss09}. The LABOCA ECDF-S Submillimetre Survey (LESS) therefore encompasses the entire region mapped by AzTEC at 1.1\,mm and reaches a uniform complementary depth of $1\sigma\approx1.2$\,mJy\,beam$^{-1}$ at 870\,\micron.  A detailed comparison involving a joint analysis of the two data-sets by a combined group from the two surveys is in progress. In this Section we carry out a preliminary comparison of the sub-mm/mm sources and their number counts in order to check for general consistency between these two data-sets.

\begin{figure}
\begin{center}
\includegraphics{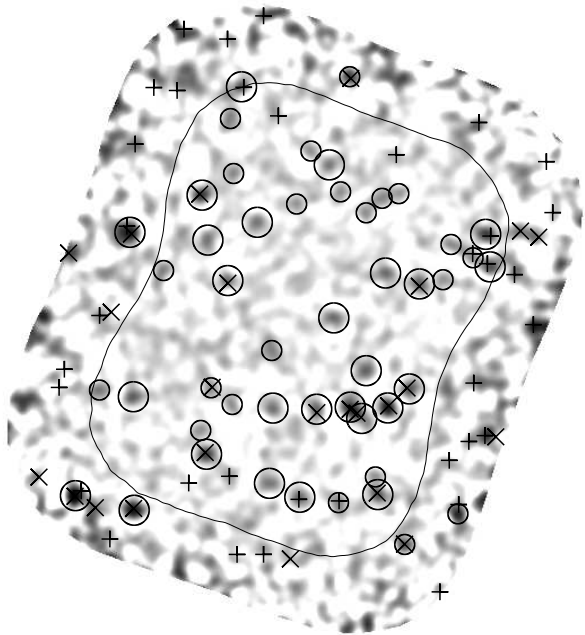}
\caption{The AzTEC/GOODS-S flux density map with the positions of AzTEC and LABOCA source candidates indicated. To distinguish between higher- and lower-significance detections, AzTEC sources detected with $S/N\ge5.0$ are shown by larger circles, while AzTEC sources detected with $S/N<5.0$ are shown by smaller circles. LABOCA sources detected with $S/N\ge5.0$ are indicated by crosses, and LABOCA sources detected with $S/N<5.0$ are indicated by pluses. The contour outlines the 50\% uniform coverage region for the AzTEC/GOODS-S survey.}
\label{fig:laboca_aztec_scmp}
\end{center}
\end{figure}

\subsection{AzTEC- and LABOCA-Detected Sources}
\label{ssec:aztec_laboca_smatch}

In Figure~\ref{fig:laboca_aztec_scmp}, we show the AzTEC 1.1\,mm map with the locations of the AzTEC and LABOCA sources. Of the 126 870\,\micron~source candidates detected with $S/N\ge3.7$ in the LABOCA catalogue (where five false detections are expected), 20 are located within the 50\% uniform coverage region of the AzTEC survey. Using a conservative search radius of 30\arcsec~(roughly the FWHM resolution of both surveys), we find that $16/20$ LABOCA sources are also detected by AzTEC with $S/N\ge3.5$. Turning this around, $16/41$ AzTEC source candidates are detected by LABOCA with $S/N\ge3.7$. Additionally, all seven of the $S/N\ge4.5$ AzTEC source candidates located in the extended (i.e. noisier) region of our map are also detected by LABOCA. AzTEC sources with probable 870\,\micron~counterparts from \citet{weiss09} are noted in Table~\ref{table:sources}. AzTEC/GS42 and AzTEC/GS43 each coincide with pairs of LABOCA source candidates separated by $<25\arcsec$.

Verifying that the degree of overlap between the two source catalogues is consistent with expectations given the completeness of the surveys is not straightforward given the different methods used to extract sources and quantify the map properties for these different surveys. This will be fully addressed in the joint analysis of both data-sets. The primary reasons for the lower fraction of AzTEC sources detected by LABOCA are 1) we use a lower $S/N$ threshold than that used for the LABOCA catalogue, and 2) the AzTEC map (scaled in flux density assuming $S_{870\mu m}/S_{1.1\rm{mm}}=2$) is slightly deeper than the LABOCA map. We stress that we do not expect a large fraction of the AzTEC-identified sources to be random noise peaks.

For 21 sources that are listed in both the LABOCA and AzTEC source catalogues, we compute the 870\,\micron~to 1.1\,mm flux ratio from the deboosted fluxes (we exclude AzTEC/GS42 and AzTEC/GS43 since each are resolved into two LABOCA sources). The mean flux ratio of this sample is $(S_{870\mu\rm{m}}/S_{1.1\rm{mm}})=2.0\pm0.6$. Scaling from 870\,\micron~to 850\,\micron~assuming a spectral index of $\beta=1.5$ ($S_{\nu}\propto\nu^{3.5}$), this corresponds to a flux ratio of $(S_{850\mu\rm{m}}/S_{1.1\rm{mm}})=2.1\pm0.6$, consistent with previous estimates from \citet{chapin09} and from scaling the 850\,\micron~and 1.1\,mm number counts derived from other surveys \citep{perera08,austermann10}. The observed flux ratios of the AzTEC/LABOCA sources are also consistent with expectations for cold ($T_d=25-50$\,K), dusty galaxies at $z\sim1-3$. From this direct source comparison there appears to be no significant systematic differences (e.g. calibration) between these two data-sets. Note however that since this estimate of the average 870\,\micron~to 1.1\,mm flux ratio is calculated solely from LABOCA-detected AzTEC sources, it may be biased high if there is a significant population of very high-redshift (or very cold) dusty galaxies that are faint at 870\,\micron.

\subsection{870\,\micron~and 1.1\,mm Number Counts}
\label{ssec:laboca_nc}

\citet{weiss09} carry out a $P(d)$ analysis on this data-set assuming four different parametrised models. Since all of the best-fit models give nearly the same results for the magnitude and shape of the number counts, we consider only the Schechter function parametrisation for the purposes of comparing the \citet{weiss09} results to ours. We note that since the LESS field covers a much larger area than the AzTEC/GOODS-S field, we might expect some differences in the source counts from these two data-sets owing to cosmic variance. The best-fit Schechter function to the LABOCA/ECDF-S data, scaled to 1.1\,mm assuming a flux ratio of $(S_{850\mu\rm{m}}/S_{1.1\rm{mm}})=2.0$ (as measured from the direct source comparison in Section~\ref{ssec:aztec_laboca_smatch}), is shown in Figures~\ref{fig:dnc} and \ref{fig:inc}. The number counts from the LABOCA/ECDF-S field derived using the semi-Bayesian method of \citet{coppin06} are also included in these figures.\footnote{ The Bayesian-derived number counts from LESS shown in Figures~\ref{fig:dnc} and \ref{fig:inc} were provided by A. Wei{\ss}.}

Considering first the results from the fluctuation analyses, we see fairly good agreement between the number counts derived from the LESS and AzTEC/GOODS-S surveys until $S_{1.1\rm{mm}}\lesssim1.0$\,mJy, at which point the LESS number counts rise more steeply. This is largely due to the differences in the analyses: while we fix $\alpha=-2.0$, \citet{weiss09} fit for this parameters and find a best-fit value of $\alpha=-3.7$ (as defined by Equation~\ref{eqn:schechter}). As explained in Section~\ref{ssec:fluc_analysis}, we choose to fix $\alpha$ to a value consistent with that measured by previous AzTEC surveys because the GOODS-S data cannot well-constrain a 3-parameter model. It is likely that the larger LABOCA/ECDF-S survey is able to put useful constraints on $\alpha$; however \citet{weiss09} do not include error estimates on their best-fit parameters, so it is not possible to quantitatively access the degree to which our results are consistent. We note that the value of $\alpha$ determined from the LESS fluctuation analysis is well-matched to our best-fit spectral index from a fluctuation analysis of AzTEC/GOODS-S assuming a single-power law model.

Comparing the Bayesian-extracted number counts from the AzTEC/GOODS-S and LESS surveys, we find that they are in excellent agreement for $S_{1.1\rm{mm}}\gtrsim3$\,mJy; however at $S_{1.1\rm{mm}}\lesssim3$\,mJy, the Bayesian-extracted number counts from the LABOCA data are significantly lower than our estimates. \citet{weiss09} noted a similar difference between their $P(d)$ and Bayesian-estimated counts and speculated that this discrepancy at low fluxes arises from source blending, since two or more faint sources blended together by the large beam would be counted as a single bright source and result in an underestimate of the number counts of SMGs at low flux densities. For this reason \citet{weiss09} place more confidence in their $P(d)$ results. However we have shown through simulations that the $P(d)$ method can be highly model dependent. As described in Section~\ref{ssec:bayes}, our simulations indicate that we largely account for confusion effects in our Bayesian-extracted number counts and that they are not significantly biased low (high) at faint (bright) flux densities due to source blending in the catalogue. We thus place more confidence in our Bayesian-derived results. Since \citet{weiss09} do not present similar tests of either of their number counts extraction algorithms, it is not possible to know the underlying cause of the discrepancies between their two different methods.

\citet{weiss09} report that the 870\,\micron~number counts from LESS are a factor of $\sim2$ lower than those from the 850\,\micron~SCUBA/SHADES survey at flux densities $S_{870\mu\rm{m}}>3$\,mJy, and note that this seems consistent with the reported under-densities in the ECDF-S of other high-redshift galaxy populations. On the contrary, we find that the 1.1\,mm number counts from AzTEC/GOODS-S are completely consistent with those measured in the AzTEC/SHADES survey. Since sub-mm/mm-selected galaxies potentially span a much larger volume of space than high-redshift galaxies selected by other photometric criterion, we would not necessarily expect to see an under-density of SMGs in this field. As the number counts derived from the LESS, AzTEC/GOODS-S, and AzTEC/SHADES fields are largely consistent assuming a reasonable scaling for the differences in observed wavelength, it seems unlikely that cosmic variance alone can account for the factor of $\sim2$ difference in the number counts derived from LESS and SCUBA/SHADES. Considering also the physically unrealistic high flux-scaling factor required to bring the SCUBA/SHADES and AzTEC/SHADES number counts into agreement ($R=2.5$), one possibility is that this discrepancy arises from systematics in the SCUBA/SHADES calibration and/or number counts extraction as discussed by \citet{austermann10}; this will be properly addressed via a source to source comparison of AzTEC and SCUBA sources in the SHADES fields (Negrello et al., in prep.).

\section{Contribution of Different Galaxy Populations to the Cosmic Infrared Background at $1.1\,\rm{mm}$}
\label{sec:cib}

The FIRAS instrument on the Cosmic Background Explorer (COBE) detected a uniform cosmic infrared background (CIRB), which peaks around 200\,\micron~\citep{puget96,fixsen98}. The total power in the infrared background is $1-2.6$ times larger than that of the optical/UV background, suggesting that the majority of radiation produced by young stars and AGNs over the cosmic history is absorbed and reprocessed by dust \citep[e.g.][]{gispert00}. If dusty starburst galaxies in the early universe account for much of the CIRB, the AzTEC/GOODS-S survey is expected to resolve a significant fraction of the CIRB.

Summing the deboosted $1.1\,\rm{mm}$ flux densities of the 41 $S/N\ge3.5$ source candidates in the AzTEC/GOODS-S map, we measure an integrated flux of $1.5\pm0.2\,\rm{Jy\,deg}^{-2}$ over the $0.075\,\rm{deg}^2$ field. Comparing this to the total energy density in the CIRB at $1.1\,\rm{mm}$ of $18-24\,\rm{Jy\,deg}^{-2}$ \citep{puget96, fixsen98}, we have resolved only $6-8$\% of the CIRB into individual galaxies. However, if we instead integrate the best-fit Schechter function model to the GOODS-S number counts down to $0\,\rm{mJy}$, we estimate that we have statistically detected $6.3\,\rm{Jy\,deg}^{-2}$, or $26-35$\%, of the CIRB at $1.1\,\rm{mm}$ with our AzTEC/GOODS-S survey. The fact that we do not resolve 100\% of the CIRB through this integration implies that simply extrapolating the Schechter function model that best fits our data at $0.5\,\mbox{mJy}<S_{1.1\rm{mm}}<6.5\,\mbox{mJy}$ to much lower flux densities significantly underestimates the number counts at the faint end. On the other hand, integrating the best-fit single power-law model to the GOODS-S number counts down to $S_{1.1\rm{mm}}=0.04$\, mJy would account for 100\% of the CIRB at 1.1\,mm; however, we know this model to be a poor description of the number counts of bright SMGs as well as physically unfeasible since it approaches infinity as $S_{1.1\rm{mm}}$ goes to zero.

We next use a stacking analysis to estimate the fraction of the CIRB at $1.1\,\rm{mm}$ that is resolved by the entire $1.4\,\rm{GHz}$ radio population. Stacking at the locations of $N=222$ radio sources in this field (up slightly from 219 radio sources used in the stacking analysis in Section~\ref{ssec:astrometry}, since we have shifted our AzTEC map to correct the astrometry), we calculate an average $1.1\,\rm{mm}$ flux density of $S_{1.1\rm{mm},radio}=660\pm78\,\mu\rm{Jy}$. Assuming that each of the radio sources distributed over an area of $A=0.075\,\rm{deg}^2$ has a flux density of $S_{1.1\rm{mm},radio}$, the radio population has a total integrated flux of $N\cdot S_{1.1\rm{mm},radio}/A=1.9\,\rm{Jy\,deg}^{-2}$, or $8-11$\% of the CIRB at $1.1\,\rm{mm}$

We finally estimate the contribution to the CIRB at $1.1\,\rm{mm}$ from MIPS 24\,\micron-selected sources using a similar stacking analysis. The average flux density from 1185 24\,\micron~sources distributed over a $0.068\,\rm{deg}^2$ area is $S_{1.1\rm{mm},24\mu\rm{m}}=290\pm26\,\mu\rm{Jy}$. The total integrated flux from 24\,\micron~sources at $1.1\,\rm{mm}$ is $5.0\,\rm{Jy\,deg}^{-2}$, or $21-28$\% of the total CIRB. This is similar to the fraction of the CIRB at 850\,\micron~resolved by 24\,\micron~sources ($29-37$\%) in the SCUBA/GOODS-N field \citep{wang06}. In contrast, a stacking analysis of 24\,\micron~sources with the BLAST maps of GOODS-S at 250, 350, and 500\,\micron~suggests that the full intensity of the CIRB at these shorter wavelengths is resolved by sources selected at 24\,\micron~\citep{devlin09,marsden09}. This possibly demonstrates the existence of a significant population of higher-redshift ($z\gtrsim3$) dust-obscured galaxies that are (statistically speaking) missed by current $\lambda\lesssim500$\,\micron~surveys, but account for $\approx2/3$ of the CIRB at longer wavelengths.

\section{Conclusions}
\label{sec:con} 

We imaged a $270\,\rm{arcmin}^2$ field towards the GOODS-S region to a confusion-limited depth of $1\sigma\approx0.6\,\rm{mJy}$ using the AzTEC camera on the ASTE, making this the deepest survey carried out to date at $1.1\,\rm{mm}$. We detect 41 SMG candidates with $S/N\ge3.5$, where roughly two are expected to be false positives arising from noise peaks. This survey is 50\% complete at $2.1\,\rm{mJy}$ and 95\% complete at $3.5\,\rm{mJy}$. We have demonstrated that the presence of confusion noise has significant consequences for the properties of the map (and hence the number counts) and must be considered when accessing the survey completeness and expected number of false detections in the source catalogue. From realistic simulations of the SMG population in this field, we estimate that $\approx26$\% of the source candidates identified in the AzTEC/GOODS-S map are actually two or more mm-bright galaxies with comparable flux densities blended together due to the low angular resolution of the ASTE beam.

We have used two very different methods to estimate the SMG number counts in this field: a fluctuation analysis where we model the distribution of flux density values in the map, and a semi-Bayesian technique where the number counts are determined from sampling the PFDs from the catalogue of SMGs. We have demonstrated that both methods are able to retrieve the correct number counts distribution from fully simulated data-sets. Furthermore, the best-fit number counts to the GOODS-S field using these different methods are consistent. We note that while our fluctuation analysis can provide good constraints on the number counts, the results depend strongly on the assumed model for the number counts distribution. The depth and large survey area of the AzTEC/GOODS-S map have resulted in the tightest constraints to date on the SMG number counts at $S_{1.1\rm{mm}}=0.5\,\rm{mJy}$. Comparing our cumulative number counts to those from several SCUBA lensing cluster surveys at 850\,\micron, the lensing cluster number counts appear to be biased high assuming reasonable values for the flux-scaling from 850\,\micron~to $1.1\,\rm{mm}$. These results are consistent with those from large-area blank-field surveys at 850\,\micron~with SCUBA, where the number counts at $S_{850\mu\rm{m}}\gtrsim2\,\rm{mJy}$ from lensing cluster surveys are systematically higher; this is most likely due to flux boosting effects not being fully treated in the number counts extraction from the lensing cluster surveys. We find that the number counts from the AzTEC/GOODS-S field are consistent with those from the $0.67\,\rm{deg}^2$ AzTEC/SHADES survey.

Comparing our source catalogue to the 870\,\micron~source list from the LABOCA/ECDF-S survey, 16/20 (80\%) LABOCA sources located within the 50\% uniform coverage region of the AzTEC/GOODS-S survey are also detected with $S/N\ge3.5$ in the AzTEC map; however since the AzTEC map is slightly deeper, only 16/41 (39\%) of the AzTEC sources in this field are listed in the LABOCA source catalogue. In contrast to the results from the LABOCA survey, we find no apparent underdensity of SMGs in this field compared to previous submm/mm surveys. For the SMGs that are listed in both catalogues, the mean 870\,\micron~to 1.1\,mm flux ratio is $2.0\pm0.6$ and is consistent with our expectations given typical dust SEDs and redshifts for these galaxies. Scaling the 870\,\micron~number counts to 1.1\,mm, we find fairly good agreement between the counts for $S_{1.1\rm{mm}}\gtrsim1.0$\,mJy ($\gtrsim3.0$\,mJy) extracted using a fluctuation analysis (Bayesian method) on each of the two surveys. We think that the discrepancies between the AzTEC and LABOCA source counts at fainter flux densities are due to differences in the assumed models used for the fluctuation analyses, and additional biases in the Bayesian method for the LABOCA analysis as noted by \citet{weiss09}. A combined analysis of these two confusion-limited surveys at 870\,\micron~and 1.1\,mm is in progress.

We resolve only $6-8$\% of the CIRB at $1.1\,\rm{mm}$ into individual mm-bright galaxies. While the 24\,\micron~population can account for the full energy density in the CIRB at $250-500$\,\micron, we estimate that 24\,\micron~sources statistically detect only $21-28$\% of the CIRB at $1.1\,\rm{mm}$, demonstrating that a population of faint dust-obscured galaxies at $z\gtrsim3$ that are largely missed at shorter wavelengths potentially contribute significantly to the total energy density in the CIRB at $1.1\,\rm{mm}$.


\section*{acknowledgements}
The authors would like to thank the anonymous referee for his/her comments and suggestions, which have greatly improved this paper. We thank N. Miller for providing us with the VLA/ECDF-S map prior to its publication, and M. Dickinson and R.-R. Chary for use of the $Spitzer$ GOODS-S IRAC/MIPS catalogues. We would also like to thank everyone who helped staff and support the AzTEC/ASTE 2007 operations, including K. Tanaka, M. Tashiro, K. Nakanishi, T. Tsukagoshi, M. Uehara, S. Doyle, P. Horner, J. Cortes, J. Karakla, and G. Wallace. The ASTE project is driven by the Nobeyama Radio Observatory (NRO), a branch of the National Astronomical Observatory of Japan (NAOJ), in collaboration with the University of Chile and Japanese institutions including the University of Tokyo, Nagoya University, Osaka Prefecture University, Ibaraki University, and Hokkaido University. Observations with ASTE were in part carried out remotely from Japan using NTT's GEMnet2 and its partner R\&E networks, which are based on the AccessNova collaboration of the University of Chile, NTT Laboratories, and the NAOJ. Support for this work was provided in part by NSF grant AST 05-40852 and a grant from the Korea Science \& Engineering Foundation (KOSEF) under a cooperative Astrophysical Research Center of the Structure and Evolution of the Cosmos (ARCSEC). This study was supported in part by the MEXT Grant-in-Aid for Specially Promoted Research (No.~20001003). KSS was supported in part through the NASA GSFC Cooperative Agreement NNG04G155A. BH and YT were financially supported by a Research Fellowship from the JSPS for Young Scientists. IA and DHH acknowledge partial support by CONACyT from research grants 39953-F and 39548-F.


\bibliography{references}

\end{document}